\documentclass[aps,pra,twocolumn,floatfix,superscriptaddress,longbibliography,showpacs,nofootinbib]{revtex4-1}

\usepackage[utf8]{inputenc}
\usepackage{graphicx}
\usepackage{dcolumn}
\usepackage{bm}
\usepackage{amsfonts}
\usepackage{amsmath}
\usepackage{amssymb}
\usepackage{braket}
\usepackage{color,soul}
\usepackage{wasysym}
\usepackage{mathrsfs}
\usepackage{times}
\usepackage[dvipsnames,svgnames,table]{xcolor}
\usepackage{ulem}
\usepackage{hyperref}
\usepackage{fancyhdr}
\usepackage[english]{babel}
\usepackage{longtable}
\usepackage{multirow}
\usepackage{float}

\hypersetup{
    unicode=true,				
    pdftoolbar=true,			
    pdfmenubar=true,			
    pdffitwindow=false,			
    pdfstartview={FitH},		
    pdfauthor={FDR},			
    colorlinks=true,			
    linkcolor=NavyBlue,			
    citecolor=Maroon,			
    filecolor=NavyBlue,			
    urlcolor=NavyBlue       	
}

\newcommand{\tf}[1]{\mathrm{#1}}

\newcommand{\td}[2]{\frac{\tf{d}{#1}}{\tf{d}{#2}}}
\newcommand{\pd}[2]{\frac{\partial {#1}}{\partial {#2}}}

\begin{document}

\title{Role of coherence in quantum-dot-based nanomachines within the Coulomb blockade regime}

\author{Federico D. Ribetto}
\affiliation{Instituto de F\'isica Enrique Gaviola (CONICET) and FaMAF, Universidad Nacional de C\'ordoba, Argentina}
\affiliation{Departamento de F\'isica, Universidad Nacional de R\'io Cuarto, Ruta 36, Km 601, 5800 R\'io Cuarto, Argentina}

\author{Ra\'ul A. Bustos-Mar\'un}\email{Corresponding author: rbustos@famaf.unc.edu.ar}
\affiliation{Instituto de F\'isica Enrique Gaviola (CONICET) and FaMAF, Universidad Nacional de C\'ordoba, Argentina}
\affiliation{Facultad de Ciencias Qu\'imicas, Universidad Nacional de C\'ordoba, Argentina}

\author{Hern\'an L. Calvo}
\affiliation{Instituto de F\'isica Enrique Gaviola (CONICET) and FaMAF, Universidad Nacional de C\'ordoba, Argentina}
\affiliation{Departamento de F\'isica, Universidad Nacional de R\'io Cuarto, Ruta 36, Km 601, 5800 R\'io Cuarto, Argentina}

\begin{abstract}
During the last decades, quantum dots within the Coulomb blockade regime of transport have been proposed as essential building blocks for a wide variety of nanomachines. This includes thermoelectric devices, quantum shuttles, quantum pumps, and even quantum motors.
However, in this regime, the role of quantum mechanics is commonly limited to provide energy quantization while the working principle of the devices is ultimately the same as their classic counterparts. Here, we study quantum-dot-based nanomachines in the Coulomb blockade regime, but in a configuration where the coherent superpositions of the dots' states plays a crucial role. We show that the studied system can be used as the basis for different forms of ``true'' quantum machines that should only work in the presence of these coherent superpositions.
We analyze the efficiency of these machines against different nonequilibrium sources (bias voltage, temperature gradient, and external driving) and the factors that limit it, including decoherence and the role of the different orders appearing in the adiabatic expansion of the charge/heat currents.
\end{abstract}

\maketitle

\section{Introduction}
\label{sec:intro}

The high degree of control and the discrete energy spectrum of coupled quantum dots (QDs), sometimes referred to as quantum dot molecules, make them especially suitable for the manipulation of charge and energy fluxes in the nanoscale. This is crucial for nanoscopic heat and charge management, the development of new quantum information technologies, and the design of different forms of quantum machines~\cite{brandes2005,zimbovskaya2013,wu2014,benenti2017,whitney2018,bustos2019}.
In this regard, experimental and theoretical studies have shown that quantum-dot-based designs may provide remarkable performances in thermoelectric devices that exchange electrical and thermal energies~\cite{prance2009,josefsson2018,beenakker1992,haupt2013,sothmann2014}.
The pumping of charge and heat on quantum-dot-based driven systems has been extensively studied~\cite{brouwer1998,watson2003,cota2005,juergens2013,haupt2013,perroni2014,taguchi2016,romero2017,terrenalonso2019,cangemi2020,zimbovskaya2020,chorley2012,roche2013,benyamini2014}.
In recent years, the reverse process in which heat or charge currents are used to propel a mechanical device has also gained considerable attention~\cite{dundas2009,bustos2013,fernandez2015,ludovico2016entropy,celestino2016,fernandez2017,calvo2017,bruch2018,hopjan2018,ludovico2018,lin2019,fernandez2019,quin2020}.

In all the above systems, it is assumed that the typical size of the device is smaller than the characteristic coherence length of the electrons. It is clear then that quantum mechanics becomes crucial for the description of these forms of nanodevices, which can be put together under the generic name of quantum machines.
Depending on the type of energy conversion involved they are usually referred to as (adiabatic) quantum motors, (adiabatic) quantum pumps (or generators), quantum heat engines, or quantum heat pumps~\cite{ludovico2016,bustos2019}.
In a quantum motor, a dc electric current is transformed into mechanical work while in a quantum pump, an ac electrical or mechanical driving is turned into a dc electric current. Quantum heat engines and heat pumps are very similar systems but the power source involves temperature gradients instead of bias voltages, and the focus is shifted from charge currents to heat currents.
\begin{figure}[h]
  \centering
  \includegraphics[width=\columnwidth]{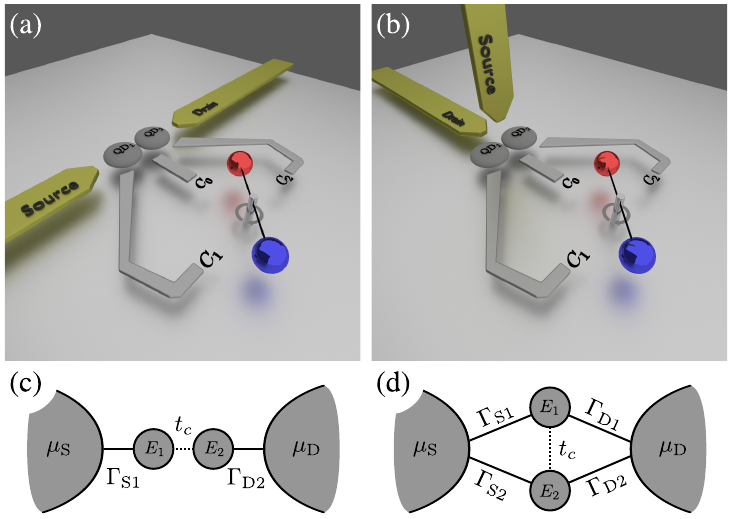}
  \caption{Example of the types of studied systems: A double quantum dot, either in series (a) or in parallel (b), coupled to some mechanical degree of freedom. Here, the dots are weakly tunnel coupled to each other and to source/drain leads (golden contacts). The dots' energies are modulated by the gate voltages generated from the \textit{capacitive} coupling to a charged rotor (silver contacts). No tunnel coupling between the dots and the rotor is considered in this model. Panels (c) and (d) are simplified schemes for the DQD system in series and in parallel, respectively. The coupling to the mechanical degree of freedom enters through the dots' energies $E_1$ and $E_2$. Lead-dot tunneling events are characterized by four tunneling rates ($\Gamma_{\tf{S}1}$, $\Gamma_{\tf{S}2}$, $\Gamma_{\tf{D}1}$, and $\Gamma_{\tf{D}2}$) while the interdot \textit{tunnel} coupling is described by $t_c$. Throughout this work, the parallel DQD without interdot coupling will be dubbed the \textit{decoupled parallel} configuration.}
\label{fig:scheme}
\end{figure}

The role of quantum mechanics on quantum-dot-based machines strongly depends on the system's conditions. Here, we focus on the \textit{adiabatic} regime, where the modulation of the system's parameters is slow as compared with the typical time spent by the electrons inside it.
However, even within this condition there are different transport regimes that should be distinguished.
For example, in the ballistic regime, described by a mean-field approximation of the electron repulsion, the working principle of adiabatic quantum pumps and motors can be attributed to interference effects of the electrons passing through the modulation region~\cite{brouwer1998,avron2004,cohen2005,bustos2013,brandner2020}.
On the other hand, in the Coulomb blockade regime, quantum effects are typically restricted to energy quantization so that the internal pumping mechanism, beyond the quantization of the transported charge, resembles that of a classical pump.
In this case, some form of rate equation relating the occupation probabilities of the quantum-dot states is typically used to describe the system, while the coherent superpositions between them can be disregarded in a first approximation.
Quantum pumping~\cite{splettstoesser2006,winkler2009,cavaliere2009,calvo2012,juergens2013,haupt2013,celestino2016,chorley2012,roche2013,benyamini2014}, shuttle transport~\cite{shekhter2007,quin2020}, and even adiabatic quantum motors~\cite{calvo2017} have been studied by using this approach.
Other strategies, like the nonequilibrium Green's function formalism, have also been used in the past to study quantum pumping within the Coulomb blockade regime~\cite{hernandez2009,deus2012}. However, the working principle of the device can also be explained by relying on a classical analog.

Based on the above, it is fair to wonder, once in the adiabatic and Coulomb blockade regimes, how ``quantum'' a nanomachine based on quantum dots can be. In this context, the weak interdot \textit{tunnel} coupling regime provides a useful platform to test the role of coherent superpositions of the dots' states. This is so because the degeneracy of dots' states brings together both occupations and coherences~\footnote{Throughout this work we will refer to the quantum superposition of the double quantum dot eigenstates as \textit{coherences}, which will be accounted for by the off-diagonal elements of the dots' reduced density matrix (see Sec.~\ref{subsec:mastereq}).} on the same time scale, which correspond to the diagonal and off-diagonal elements of the dots' reduced density matrix, respectively.
As a consequence of that, coherences survive even at the steady state of the system.
In particular, in Ref.~[\onlinecite{riwar2010}] charge pumping was studied for a double quantum dot (DQD) coupled in series.
They found that the coupling between coherences and occupations is responsible for charge pumping.
However, given that both coherences and electron transport rely entirely on the interdot coupling, taking this coupling to zero trivially sets the current to zero. Thus, although the ``quantumness'' of the pumping mechanism is clearly present, its effect is somewhat hidden.
On the other hand, in Ref.~[\onlinecite{hiltscher2010}] the authors analyzed charge pumping in an Aharonov-Bohm interferometer configuration of the dots. As in this case there is no explicit interdot coupling, the role of quantum superposition becomes more clear. Similarly, for quantum systems weakly coupled to thermal reservoirs, the role of coherences in the thermodynamics was analyzed in Ref.~[\onlinecite{cuetara2016}] for degenerate quantum dots, while the relevance of coherent effects in the adiabatic dynamics is discussed in Ref.~[\onlinecite{bhandari2021}].

In this work, we exploit the weak interdot coupling regime in a DQD to analyze the role of coherence in a broad class of quantum machines such as charge/heat pumps and nanomotors driven by bias voltages or temperature gradients. We use a real-time diagrammatic approach~\cite{koenig1996,splettstoesser2006}, that takes into account both the occupations and coherences in lowest order in the tunnel coupling to the leads. Importantly, the inclusion of the off-diagonal elements of the reduced density matrix goes beyond what is understood by the ``sequential tunneling regime''.

We start our description of the known~\cite{riwar2010} charge pumping mechanism for the DQD in series and then we focus on the parallel configuration, which highlights the role of quantum superposition in the steady state of quantum transport. These devices are coupled to some classical degree of freedom, which provides the necessary dot energy level modulation for the machine to become operational. Figure~\ref{fig:scheme} illustrates the considered models for our proposal. We show that the above-mentioned regime dominated by coherences also applies to quantum motors fueled by a finite bias voltage. 
In particular, we demonstrate that the existence of coherences (and their coupling to the occupations) is always necessary for these devices to function. In addition, we include in our description an external force acting on the (classical) mechanical component of the system. Such a force allows us to bring together the two operation modes (pump and motor) of the device on a same basis. These ideas are also extended to the case where the leads are subjected to different temperatures, giving rise to quantum heat engines and refrigerators enabled by coherence.
We analyze the performance of these machines and the factors that limit it. This includes (1) a decoherence model that destroys the coherent superpositions of the DQD states, by reducing the contribution of the off-diagonal elements of the reduced density matrix in the transport properties, and (2) the role of the different orders of the adiabatic expansion of the involved currents, which leads to leaking effects.

The paper is organized as follows. In Sec.~\ref{sec:model}, we present the theoretical framework, including a brief overview of the real-time diagrammatic approach, the expressions for the observables, the definitions of the efficiencies, and the used decoherence model. In Sec.~\ref{sec:dqd}, we apply the formalism to the particular example of a DQD weakly tunnel coupled to two external leads and capacitively coupled to a rotor. In Sec.~\ref{sec:example}, we study the performance of the different operational regimes of the decoupled parallel configuration of the DQD. Finally, in Sec.~\ref{sec:conclusions}, we summarize the main results.

\section{Theoretical framework}
\label{sec:model}

\subsection{Hamiltonian model}

We consider a system composed of QDs in which mechanical and electronic degrees of freedom are present and coupled to each other. From now on we call this system the \textit{local system}, and we model it by the following Hamiltonian:
\begin{equation}
  \hat{H}_\tf{local} = \hat{H}_\tf{el}(\hat{\bm{X}})+\frac{\hat{\bm{P}}^2}{2m} + U_\tf{ext}(\hat{\bm{X}},t),  
\label{eq:hlocal}
\end{equation}
where $\hat{\bm{X}} = (\hat{X}_1,..., \hat{X}_N)$ is the vector (operator) of mechanical coordinates while $\hat{\bm{P}} = (\hat{P}_1,...,\hat{P}_N)$ represents their associated momenta, $m$ is the effective mass related to $\hat{\bm{X}}$, and $U_\tf{ext}$ denotes some external mechanical potential that may be acting on the local system.
We use an explicit time dependence in $U_\tf{ext}$ to denote that an external and nonconservative force might be acting on the mechanical subsystem (see below).
The Hamiltonian $\hat{H}_\tf{el}$ includes the electronic degrees of freedom of the system, that are participating in the transport, as well as their coupling to the mechanical ones through
\begin{equation}
  \hat{H}_\tf{el}(\hat{\bm{X}}) = \sum_i E_i(\hat{\bm{X}}) \ket{i}\!\bra{i},
\label{eq:hel}
\end{equation}
where the sum runs over all possible electronic many-body eigenstates $\ket{i}$.
The system is then weakly coupled to external leads so the \textit{total} Hamiltonian reads
\begin{equation}
  \hat{H}_\tf{total} = \hat{H}_\tf{local} + \sum_r \hat{H}_r + \hat{H}_\tf{tun}. 
\end{equation}
The leads are described as noninteracting electron reservoirs through the Hamiltonian
\begin{equation}
  \hat{H}_r = \sum_{k\sigma} \epsilon_{rk} \hat{c}_{rk\sigma}^\dag \hat{c}_{rk\sigma},
\end{equation}
where $\hat{c}_{rk\sigma}^\dag$ ($\hat{c}_{rk\sigma}$) is the creation (annihilation) operator for an electron with state index $k$ and spin projection $\sigma = \{\uparrow,\downarrow\}$ in the lead $r$, which we typically take as source and drain reservoirs, i.e., $r = \{\tf{S},\tf{D}\}$. These reservoirs are assumed to be always in equilibrium, characterized by a temperature $T_r$ and an electrochemical potential $\mu_r$.~\footnote{Throughout this work, we will use the terms ``contact'', ``lead'' and ``reservoir'' interchangeably.}
Finally, the tunnel coupling between the local system and the leads is given by the tunnel Hamiltonian
\begin{equation}
  \hat{H}_\tf{tun} = \sum_{rk\sigma \ell} (t_{r\ell} \hat{d}_{\ell\sigma}^\dag \hat{c}_{rk\sigma} + \tf{H.c.}),
  \label{eq:htun}
\end{equation}
where $t_{r\ell}$ denotes the tunneling amplitude, which we assume to be $k$ and spin independent for simplicity.
The fermionic operator $\hat{d}_{\ell\sigma}^\dag$ ($\hat{d}_{\ell\sigma}$) creates (annihilates) an electron with spin $\sigma$ in the quantum dot $\ell$ composing the local system. The tunnel-coupling strengths, defined as $\Gamma_{r\ell} = 2 \pi \rho_r |t_{r\ell}|^2$, quantitatively describe the rate at which electrons enter (leave) the quantum dot $\ell$ from (to) the $r$ reservoir. We also define the total tunnel rate as $\Gamma = \sum_{r\ell} \Gamma_{r \ell}$.
The reservoirs are taken to be in the wideband limit where their densities of states $\rho_r$ are assumed to be energy independent.
Throughout this paper, we set $e = 1$ for the absolute value of the electron charge and $\hbar = 1$.

\subsection{Stationary state regime}
\label{subsec:stationary}

We suppose that the dynamics of the electronic and mechanical degrees of freedom are well separated from each other, and therefore we can treat them through the Born-Oppenheimer approximation. Under this approximation, the mechanical coordinates can be treated as classical variables obeying the following Langevin-like equation
\begin{equation}
  m \ddot{\bm{X}} +\bm{F}_\tf{ext} = \bm{F}+\bm{\xi},
\label{eq:langevin}
\end{equation}
where $\bm{F}= -\braket{\nabla \hat{H}_\tf{el}} = \tf{i} \braket{[\nabla \hat{H}_\tf{el},\hat{\bm{P}}]}$ is the mean value of the current-induced forces (CIFs) while $\bm{\xi}$ stands for its fluctuation. Later on we will see that a friction component arises from expanding $\bm{F}$ in terms of the velocity of the mechanical coordinates~\cite{bode2011,bustos2013,cunningham2015,gu2016,bai2016,calvo2017,lu2019,chen2019}.
The term $\bm{F}_\tf{ext}$ represents an external force applied to the mechanical part of the local system and is related to the potential $U_\tf{ext}$ in Eq.~(\ref{eq:hlocal}). This force will be, in general, opposed to the bias-induced direction of the CIF, so we define it with a minus sign for convenience.
As we shall see later, in our model such a quantity appears as the key tool to set up the different operation modes of the electromechanical device.
If we manage to calculate the expectation value of the CIF (see Sec.~\ref{subsec:obs}) then we can use Eq.~(\ref{eq:langevin}) to integrate the classical equations of motion and derive the effective dynamics of the local system, including both electronic and mechanical degrees of freedom. In realistic systems, friction and stochastic forces may have different origins, such as the coupling to other phononic degrees of freedom. Here, however, we are only interested in the quantum effects of CIFs. Thus, we will only take into account friction and stochastic forces that arise from the coupling to the electronic degrees of freedom.

Before continuing, some comments about the system are in order. First, we will focus on systems whose mechanical part is capable of reaching a stationary regime characterized by a steady cyclic motion (with some frequency $\Omega \propto \dot{\bm{X}}$) and whose dynamics can be described by an \textit{angular} Langevin equation.
If we assume that this \textit{rotor} follows a circular trajectory then only one parameter, the angle $\theta$, is needed for the study of its dynamics. In this case we can project Eq.~(\ref{eq:langevin}) on the angular direction $\hat{\bm{\theta}}$ to obtain the following angular form~\cite{calvo2017,bustos2019}:
\begin{equation}
  \ddot{\theta} = \frac{1}{\mathcal{I}} (\mathcal{F}-\mathcal{F}_\tf{ext} + \xi_\theta),
\label{eq:anglangevin}
\end{equation}
where $\mathcal{I}$ is the moment of inertia of the mechanical subsystem, $\mathcal{F}$ is the current-induced torque, $\mathcal{F}_\tf{ext}$ is the torque associated with the external force, and $\xi_{\theta}$ is the stochastic torque which comes from the angular projection of the CIF's fluctuation.

Second, in addition to the assumption of cyclic mechanical motion, we consider that the terminal velocity reached by the system is constant, i.e., $\dot{\theta} = \Omega$, during the whole cycle.
This is also justified for large values of $\mathcal{I}$, where the variation of the angular velocity (together with its fluctuations) along the cycle becomes negligible~\cite{fernandez2015,fernandez2017,calvo2017,bustos2019}.
Both numerical and analytical procedures for the calculation of $\dot{\theta}$, before and after reaching stationarity, have been carried out in 
Refs.~[\onlinecite{calvo2017}] and [\onlinecite{bustos2019}].

We are now in position to derive a relation between the work related to the torques $\mathcal{F}$ and $\mathcal{F}_\tf{ext}$.
This is done by integrating Eq.~(\ref{eq:anglangevin}) over a whole period of the system at the stationary state, yielding~\cite{bustos2019}
\begin{equation}
  \mathcal{W}_F = \int_0^\tau \mathcal{F} \dot{\theta} \, \tf{d}t = \int_0^{\tau} \mathcal{F}_\tf{ext} \dot{\theta} \, \tf{d}t = \mathcal{W}_\tf{ext},
\label{eq:stationary}
\end{equation}
where $\tau=2\pi/\Omega$ is the period of the cycle. The equation implies that, once the cycle is completed, the work related to the CIF is balanced by the work done by the external mechanical force. This equality is fundamental in the sense that it defines the stationary state condition mentioned before and allows us to extract the value of $\dot\theta=\Omega$.

\subsection{Generalized master equation}
\label{subsec:mastereq}

In this section we introduce the formalism that describes the dynamics of the electronic part of the system. This will allow us to calculate the expectation value of the CIF, together with other relevant observables like charge and heat currents, while exactly taking into account the strong Coulomb interaction in the local system. We assume that, before certain initialization time $t_0$, the leads and the local system are decoupled, such that the total density matrix can be factorized as $\hat{\rho} = \hat{p}_\tf{res} \otimes \hat{p}$. Here $\hat{p}_\tf{res}$ describes
the leads' density matrix, while $\hat{p}$ represents the \textit{reduced} density matrix of the local system. When both subsystems are coupled together, the relevant information of the local system dynamics at times $t > t_0$ is encoded in $\hat{p}(t) = \tf{Tr}_\tf{res} [\hat{\rho}(t)]$, where $\tf{Tr}_\tf{res}$ is the trace over the reservoirs' degrees of freedom.
The time evolution of the matrix elements is governed by the generalized master equation~\cite{splettstoesser2006}
\begin{widetext}
\begin{equation}
  \td{}{t} p_{\alpha,\beta}(t) = \sum_{\alpha',\beta'} \left[-\tf{i} L_{\alpha,\beta}^{\alpha',\beta'}(t) p_{\alpha',\beta'}(t)
  + \int_{-\infty}^t \tf{d}t' W_{\alpha,\beta}^{\alpha',\beta'}(t,t') p_{\alpha',\beta'}(t') \right],
\label{eq:master}
\end{equation}
\end{widetext}
where $p_{\alpha,\beta}(t)= \bra{\alpha} \hat{p}(t) \ket{\beta}$ and we have taken the limit $t_0 \rightarrow -\infty$, in order to neglect any transient effect. The first term on the right-hand side of this equation takes into account the internal dynamics of the QDs through the Liouvillian superoperator $L \, \bullet \equiv [\hat{H}_\tf{el}, \bullet]$, while the second term describes state transitions due to electron tunneling processes between the leads and the local system. This is quantified by the kernel superoperator $W$~\footnote{The $W$ superoperator and related matrices $\bm{W}$ and $\bm{W}^\tf{eff}$ should not be confused with the mechanical work $\mathcal{W}_F$ or $\mathcal{W}_\tf{ext}$.} representing all irreducible diagrams in the Keldysh double contour~\cite{koenig1996}, and whose matrix element $W_{\alpha,\beta}^{\alpha',\beta'}$ describes the transition between states $\alpha'$ and $\beta'$ at time $t'$, and states $\alpha$ and $\beta$ at time $t$, due to tunnel processes.

To simplify the notation, we gather the diagonal (occupations) and off-diagonal (coherences) elements of the reduced density matrix into a vector, $\hat{p} \rightarrow \bm{p} \equiv (\bm{p}_\tf{d}, \bm{p}_\tf{n})^\tf{T}$, yielding a matrix representation for both $W$ and $L$ superoperators, i.e., $W \rightarrow \bm{W}$ and $L \rightarrow \bm{L}$.
Here, the diagonal and off-diagonal elements of the reduced density matrix are contained in $\bm{p}_\tf{d}$ and $\bm{p}_\tf{n}$, respectively. 
Thus we can think of $\bm{W}$ and $\bm{L}$ as composed by the following block matrices:
\begin{equation}
  \begin{array}{lllllll}
    \bm{W} & = & \left(\begin{array}{cc}
      \bm{W}_\tf{dd} & \bm{W}_\tf{dn}\\
      \bm{W}_\tf{nd} & \bm{W}_\tf{nn}
    \end{array}\right), \quad & \bm{L} & = & \left(\begin{array}{cc}
      \bm{L}_\tf{dd} & \bm{L}_\tf{dn}\\
      \bm{L}_\tf{nd} & \bm{L}_\tf{nn}
    \end{array}\right)
  \end{array}.
  \label{eq:kernel}
\end{equation}
As we already mentioned, Eq.~(\ref{eq:hel}) tells us that the dots' energy levels are affected by the 
cyclic mechanical motion, characterized by a frequency $\Omega$ proportional to the mechanical velocities 
$\dot{\bm{X}}$. If we assume that the dwell time of the electrons in the local system is much shorter than 
the mechanical period $\tau$, then it is possible to perform a frequency expansion on $\boldsymbol{p}(t)$~\cite{splettstoesser2006,riwar2010,juergens2013,calvo2017,bustos2019}.
Strictly speaking, this \textit{adiabatic} approximation holds if the adiabaticity condition 
$\Omega / \Gamma \ll k_\tf{B} T / \delta \epsilon$ is satisfied, where $\delta \epsilon$ stands for the 
energy amplitude of the QDs' energy levels.
This allows us to expand the reduced density matrix as $\bm{p} (t) = \sum_{k \geq 0} \bm{p}^{(k)} (t)$ 
with $\bm{p}^{(k)} \sim (\Omega / \Gamma)^{k}$. The first term, $\bm{p}^{(0)} (t)$, represents the 
steady-state solution at which the electronic part of the system arrives when the mechanical
coordinates are \textit{frozen} at time $t$. In other words, this order corresponds to the adiabatic
electronic response to the mechanical motion. Note that here we are referring to the steady state of
the electronic part of the system, which should not be confused with the steady-state regime of the
mechanical degrees of freedom mentioned in the previous section. From now on, every time we talk about 
stationarity, it will be referred to as the mechanical part of the local system. Higher-order terms ($k > 0$) 
represent nonadiabatic corrections due to retardation effects in the electronic
response mentioned earlier.

On top of this adiabatic expansion for small $\Omega$, we perform a perturbative expansion in the
tunnel coupling strengths, taking only terms up to first order in $\Gamma$ (which is reasonable in the
weak tunnel coupling limit considered here). Higher-order processes, like cotunneling, are therefore
ignored throughout this paper. This double expansion gives rise to the following hierarchy of 
equations~\cite{splettstoesser2006,cavaliere2009,riwar2010,juergens2013,calvo2017,bustos2019}:
\begin{eqnarray}
  \bm{W}^\tf{eff} \bm{p}^{(0)} & = & \bm{0}, \notag \\
  \bm{W}^\tf{eff} \bm{p}^{(k)} & = & \td{}{t} \bm{p}^{(k-1)},
  \label{eq:W^eff}
\end{eqnarray}
where we have defined the effective kernel $\bm{W}^\tf{eff}$ as the zero-frequency Laplace transform of $\bm{W}-\tf{i}\bm{L}$, 
with both matrices evaluated up to first order in $\Gamma$.
We omit the frequency order superscript in the effective kernel since at this level of approximation it is always $\mathcal{O}(\Omega^0)$.
We remark that the above order-by-order expansion relies in (1) taking up to first-order terms in $\Gamma$, and (2) that the off-diagonal elements $p_{\alpha,\beta}$ of the reduced density matrix are of the same order as the diagonal ones. The latter is due to the fact that in the weak interdot coupling regime considered in this work (see Sec.~\ref{sec:dqd}), the internal parameters of the DQD system are of the same order as $\Gamma$, and therefore we can take $\bm{L}\propto\Gamma$ in this particular case.
In other words, we will always work with the \textit{secular} elements of the reduced density matrix, defined as those $p_{\alpha,\beta}$ where $|E_\alpha-E_\beta|\lesssim\Gamma$.~\footnote{The term ``secular'' in this context refers to the involved timescales of the matrix elements of $\hat{p}$, as compared to the typical dwelling time of the electrons in the DQD system.} Nonsecular elements of $\hat{p}$ can be safely neglected in this lowest order in $\Gamma$ approximation~\cite{reckermann2010}.
It should be noted that, even at this level of the approximation, the above set of equations goes beyond the sequential tunneling approach since, for the regime considered, the secular part of $\hat{p}$ also includes those off-diagonal elements ($\alpha\ne\beta$) that cannot be disregarded.
These equations, combined with the normalization condition on the reduced density matrix,
$\bm{e}^\tf{T} \bm{p}^{(k)} = \delta_{k 0}$, allows us to iteratively calculate $\bm{p}^{(0)}$ and any
nonadiabatic correction $\bm{p}^{(k)}$.
The vector $\bm{e}^\tf{T} \equiv (1, \ldots, 1, 0, \ldots, 0)^\tf{T}$ is
a representation of the local system's trace operator, where the number of ones equals the
dimension of the reduced Hilbert space. In light of this, the nonadiabatic corrections can be written
as~\cite{cavaliere2009}
\begin{equation}
 \bm{p}^{(k)} = \left( \tilde{\bm{W}}^{-1} \td{}{t} \right)^k \bm{p}^{(0)}.
 \label{eq:pk}
\end{equation}
Here $\tilde{\bm{W}}$ represents the (invertible) pseudo kernel, defined as $\tilde{W}_{ij} \equiv 
W^\tf{eff}_{ij}-W_{ii}^\tf{eff}$ for the $\tf{dd}$ block and $\tilde{W}_{ij} \equiv W^\tf{eff}_{ij}$ for the remaining ones, in order to exclude the zero eigenvalue through the normalization condition.
Since the effective kernel is linear in $\Gamma$, the $k$ term of the reduced density matrix, $\bm{p}^{(k)}$, is
proportional to $(\Omega/\Gamma)^k$. This forces us to assume $\Omega < \Gamma$, in order to avoid any 
divergence~\cite{splettstoesser2006}. More specifically, as the time dependence considered in this work enters through the energy levels of the DQD, the above expansion leads to the aforementioned adiabaticity condition~\cite{bustos2019}.
Once we get $\bm{p}^{(0)}$ (see Appendix~\ref{app:kernel}) and any required nonadiabatic correction $\bm{p}^{(k)}$, we can proceed with the calculation of all observables related to the performance of adiabatic quantum machines. In the next section we discuss the procedure used to achieve this task.

\subsection{Observables}
\label{subsec:obs}

Now we are going to make use of the formalism described in the previous section to determine the
expectation values of a set of observables.
First, we consider the charge current $I_r (t) \equiv \braket{\hat{I}_r(t)}$ and the heat current
$J_r (t) \equiv \braket{\hat{J}_r(t)}$, both associated with the $r$ lead.
For these quantities we take the sign convention that in each lead the particle and heat currents 
are positive when particles and heat are flowing toward the lead; thus we can write the currents 
in the lead $r$ as
\begin{align}
  I_r (t) &=  \td{}{t} \tf{Tr} [\hat{N}_r \hat{\rho} (t)], \\
  J_r (t) &= \td{}{t} \tf{Tr} [(\hat{H}_r - \hat{N}_r \mu_r) \hat{\rho}(t)],
\end{align}
where $\hat{N}_r$ is the number operator for the electrons in the reservoir $r$. We also address the
CIF which, unlike the previous observables, constitutes a local quantity. As we showed before, Eq.~(\ref{eq:hel})
tells us that the mechanical part of the system only interacts with the local parameters of the dots
via their many-body eigenenergies. This implies that the CIF only consists of fermionic dot operators,
and therefore we can write its expectation value as
\begin{equation}
  \bm{F}(t) = - \underset{\tf{local}}{\tf{Tr}}[\nabla \hat{H}_\tf{el} \hat{p}(t)],
\end{equation}
where the gradient is taken with respect to the mechanical coordinates $\bm{X}$.

The adiabatic expansion developed in Sec.~\ref{subsec:mastereq} can also be performed over any observable 
$R$ of interest ($I_r$, $J_r$, and $\bm{F}$ in our case),
\begin{equation}
  R(t) = \sum_{k \geq 0} R^{(k)}(t) . 
\end{equation}
To lowest order in $\Gamma$, the $R^{(k)}$ terms can be written as
\begin{equation}
  R^{(k)} = \bm{e}^\tf{T} \bm{A}^R \bm{p}^{(k)}, 
\end{equation}
where $\bm{A}^R$ stands for the kernel/matrix associated with the observable $R$. The charge and heat currents
flowing from the lead $r$ into the device are represented by the following kernels~\cite{haupt2013,juergens2013,calvo2017,bustos2019}
\begin{align}
  [\bm{A}^{I_r}]_{ij} &= - n_i [\bm{W}^\tf{eff}_r]_{ij}, \label{eq:WIr} \\
  [\bm{A}^{J_r}]_{ij} &= - (E_i - \mu_r n_i) [\bm{W}^\tf{eff}_r]_{ij}, 
\end{align}
where $n_i$ and $E_i$ are the number of particles and energy associated with the local system's
eigenstate $\ket{i}$, respectively, and $\bm{W}^{\tf{eff}}_r$ is the $r$-lead evolution kernel
such that $\bm{W}^{\tf{eff}} = \sum_r \bm{W}^{\tf{eff}}_r$. Regarding the $\nu$ component of the CIF,
we can directly construct a diagonal matrix from
\begin{equation}
  [\bm{A}^{F_\nu}]_{ij} = - \pd{E_i}{X_\nu} \delta_{ij},
\end{equation}
where again we make use of its local condition, such that these elements do not depend on the effective kernel~\cite{calvo2017,bustos2019}.

As in the case of $\bm{p}^{(0)}$, the zeroth-order terms $I^{(0)}_r (t)$ and
$J^{(0)}_r (t)$ describe the steady-state currents flowing through the system in a stationary
situation where all time-dependent parameters are kept constant at time $t$.
The only way for these terms to be nonzero is when the system is subject to a bias voltage or a temperature gradient since, in this case, the time variation of the mechanical parameters has no effective role in the observables. Higher-order terms represent additional contributions to the steady-state currents due to the delayed response of the system to the mechanical motion.
From the above-defined currents $R = \{ I_r, J_r \}$, we will work with their integrated quantities over a modulation cycle, i.e., $Q_R^{(k)}=\int_0^\tau R^{(k)} \tf{d}t$. In particular, we will refer to $Q_R^{(1)}$ as the pumped charge/heat per cycle due to the first-order charge/heat current $R^{(1)}$.

A similar analysis applies to the CIF, where we take contributions up to first order in the mechanical velocity $\Omega$, i.e., $\bm{F}(t) = \bm{F}^{(0)} + \bm{F}^{(1)}$.
The lowest-order term can be split into (1) an equilibrium contribution, which is conservative and it can be interpreted as the Helmholtz free energy of the local system, and (2) a nonequilibrium term, which appears as a consequence of temperature gradients or bias voltages among the leads~\cite{calvo2017}.
The first adiabatic correction to the CIF, proportional to $\Omega$, gives the frictional force that dissipates energy from the mechanical part of the local system toward the electronic reservoirs~\cite{bustos2019}.
For systems with multiple mechanical degrees of freedom, it also contributes to the energy exchange between modes and, for finite voltages, it can even allow the flux of energy from the leads toward the mechanical degrees of freedom~\cite{lu2010,bode2011}.

If we now perform an adiabatic expansion of the torque $\mathcal{F}$, integrate it over a cycle, and use Eq.~(\ref{eq:stationary}), we get the 
relation
\begin{equation}
  \mathcal{W}_F = s \sum_k \left( \int_0^{2\pi} \frac{\tf{d}\theta}{k!} \left. \pd{^k \mathcal{F}}{\dot{\theta}^k} \right|_{\dot{\theta} = 0}
  \right) \dot{\theta}^k = s \sum_k \mathcal{C}_F^{(k)} \dot{\theta}^k,
  \label{eq:stat_order}
\end{equation}
where $s$ is the sign of $\dot{\theta}$ and gives the direction in which the trajectory is traversed. Here, we defined the force coefficients $\mathcal{C}_F^{(k)}$ which are independent of the direction of motion of the system, not obvious \textit{a priori}~\cite{bustos2019}. If we take terms up to $k = 1$, the angular velocity can be obtained from Eq.~(\ref{eq:stat_order}) as follows:
\begin{equation}
  \dot\theta = \Omega = \frac{\mathcal{C}_F^{(0)} - \mathcal{C}_\tf{ext}}{-\mathcal{C}_F^{(1)}},
  \label{eq:dottheta}
\end{equation}
where we defined $\mathcal{W}_\tf{ext} = s\mathcal{C}_\tf{ext}$ to keep track of every term’s sign. Note that, for the mechanical subsystem to achieve a stationary regime in the present model, the stability condition $\mathcal{C}_F^{(1)} < 0$ should be fulfilled, which implies a positive ``friction coefficient''~\cite{bustos2019}.

\subsection{Efficiency}
\label{subsec:eta}

Previously we stated that the mechanical subsystem performs a cyclic motion along a circular
trajectory while affecting the dots' energy levels. If we define a closed trajectory $\mathcal{C}$ for
the system's parameters that are being modulated, then the work $\mathcal{W}_F^{(0)}$ done by the 
zeroth-order contribution of the CIF can be calculated by performing a line integral of $\bm{F}^{(0)}$ 
along this trajectory or, for two parameters and with the aid of Stokes' theorem,
we can calculate it in the following way
\begin{equation}
  \mathcal{W}_F^{(0)} = \iint_\mathcal{S} \nabla \times \bm{F}^{(0)} \cdot \tf{d} \bm{S} \equiv 
						\iint_\mathcal{S} \bm{\mathcal{B}}^F \cdot \tf{d} \bm{S}.
\label{eq:B^F}
\end{equation}
This means that the work associated with the zeroth-order CIF can be understood as the surface integral of
a curvature vector $\bm{\mathcal{B}}^F = \nabla \times \bm{F}^{(0)}$ (the curl of the force), which is frequency independent.
Analogously, we can define a pumping curvature for the charge current flowing from/to reservoir $r$ as
\begin{equation}
  Q^{(1)}_{I_r} = \oint_\mathcal{C} \pd{I_r^{(1)}}{\dot{\bm{X}}} \cdot \tf{d}\bm{X} = \iint_\mathcal{S} \bm{\mathcal{B}}^{I_r} \cdot \tf{d}\bm{S},
  \label{eq:B^I}
\end{equation}
and the same can be done for the pumped heat $Q^{(1)}_{J_r}$, via the curvature
$\bm{\mathcal{B}}^{J_r}$. These relations (which are only valid to first order in $\Omega$) highlight the
geometrical nature of these observables in the sense that they only depend on the chosen trajectory
$\mathcal{C}$~\cite{calvo2012,juergens2013,pluecker2017}.

Equations like (\ref{eq:B^F}) and (\ref{eq:B^I}) provide a geometrical approach to the study of adiabatic quantum devices, which has been discussed by several authors~\cite{thouless1983,brouwer1998,pluecker2017,bhandari2020}.
One immediate conclusion from these equations is that the trajectory followed by the modulation parameters should enclose a finite area.
This implies that there must be at least two out-of-phase parameters modulating the device.

With the help of the geometric curvatures $\bm{\mathcal{B}}^F$, $\bm{\mathcal{B}}^{I_r}$, and $\bm{\mathcal{B}}^{J_r}$, one can set a convenient working point in the parameters' space around which a trajectory $\mathcal{C}$ will be defined. For example, if the goal is to design a nanomotor, this trajectory should enclose regions of large $\mathcal{B}^F$. On the other hand, if a charge pump is desired, then we should create a closed trajectory over regions where $\mathcal{B}^{I_r}$ is large.
All these quantities, together with their integrals, are not independent but related via order-by-order energy conservation, Onsager's reciprocal relations, and the second law of thermodynamics~\cite{ludovico2016,ludovico2016entropy,benenti2017,whitney2018,bustos2019}.

The order-by-order energy conservation is given by~\cite{juergens2013,bustos2019}
\begin{equation}
 \sum_r \left( Q_{I_r}^{(k)} \delta V_r + Q_{J_r}^{(k)} \right) = -\mathcal{W}_F^{(k-1)}. \label{eq:energycons}
\end{equation}
Here, the superscript $(k)$ indicates the order in the frequency expansion of the integrated observables, and $\delta V_r = \delta \mu_r / e$ where $\delta \mu_r = \mu_r -\mu_0$ ($\mu_0$ is the reference chemical potential).
The above equation can be useful to identify energy losses. For example, in a nanomotor driven by a bias voltage, energy losses at $k=1$ are only due to the pumped heat $Q_{J_r}^{(1)}$ resulting from the modulation of the system's parameters; see also Ref.~[\onlinecite{bustos2019}].

Onsager's reciprocal relations appear in the linear regime of transport, characterized by low bias voltages, small temperature gradients, and low velocities of the modulation parameters~\cite{bustos2013,ludovico2016}.
For example, in a two-lead configuration with $r=\{\tf{S},\tf{D}\}$ and when the leads are kept at the same temperature (i.e., $\delta T_r = 0$), Onsager's reciprocal relations imply
\begin{equation}
\tilde{Q}_{I}^{(1)} \Delta V = -\mathcal{W}_F^{(0)},
\label{eq:onsager_charge}
\end{equation}
where $\tilde{Q}_{I}^{(1)}\equiv (\tilde{Q}_{I_\tf{S}}^{(1)}-\tilde{Q}_{I_\tf{D}}^{(1)})/2$, $\delta V_{\tf{S},\tf{D}} = \pm \Delta V/2$, and we use a tilde in the pumped charge to denote that this quantity is being evaluated in the limit of zero bias. Similarly, when a temperature gradient at zero bias voltage is applied between the contacts we have
\begin{equation}
\sum_r \tilde{Q}_{J_r}^{(1)} \frac{\delta T_r}{T_r} = -\mathcal{W}_F^{(0)},
\label{eq:onsager_heat}
\end{equation}
where, again, the tilde states that the pumped heat is evaluated at equilibrium. Onsager's relations such as those shown in Eqs.~(\ref{eq:onsager_charge}) and (\ref{eq:onsager_heat}) provide a general strategy for developing novel devices from the reciprocal of known machines. For example, an adiabatic quantum motor is in essence an adiabatic quantum pump working in reverse, at least at low bias voltages~\cite{bustos2013,calvo2017}.

The second law of thermodynamics can be expressed in the following form for the type of systems treated here~\cite{bustos2019}:
\begin{equation}
 \sum_k \left[ \mathcal{W}_F^{(k)} + \sum_r \left( Q_{I_r}^{(k)} \delta V_r + Q_{J_r}^{(k)} \frac{\delta T_r}{T_r} \right) \right] \leq 0,
 \label{eq:2ndlaw}
\end{equation}
where $\delta T_r = T_r -T_0$ ($T_0$ is the temperature of reference).
Equation~(\ref{eq:2ndlaw}) allows us to derive bound expressions for the device's efficiencies which, as usual, are defined as the ratio between the output and input powers per cycle~\cite{bustos2019}.
Before doing this, we first need to know how to determine the operational mode of the device, namely, whether the device acts like a motor or a pump. 
Using Eq.~(\ref{eq:dottheta}), the sign $s$ of the constant velocity $\dot{\theta}$ can be determined and, with it, the sign of $\mathcal{W}_\tf{ext}$. The latter determines the direction of the energy flux between the local system and the external agent that is acting on it through $U_\tf{ext}$; see Eq.~(\ref{eq:hlocal}). If $\mathcal{W}_\tf{ext} > 0$, the energy current flows from the leads to the dots and, there, it is transformed into mechanical work, so the device operates as an electric motor/heat engine depending on the nonequilibrium source. On the contrary, if $\mathcal{W}_\tf{ext} < 0$ the external agent is performing mechanical work which is then dissipated through the dots to the leads, so the device operates as a pump.

Now considering that only a bias voltage is applied, the generated electrical current delivers an input
energy $Q_I \Delta V$ per cycle,~\footnote{In a source/drain setup with symmetric bias ($\delta V_{\tf{S},\tf{D}} = \pm \Delta V$) we define $I=(I_\tf{S}-I_\tf{D})/2$.} while the output energy is $W_\tf{ext}=\mathcal{W}_F^{(0)} + \mathcal{W}_F^{(1)}$. Thus, the efficiency of this \textit{electrical motor} is given by
\begin{equation}
  \eta_\tf{em} = - \frac{\mathcal{W}_F^{(0)} + \mathcal{W}_F^{(1)} }{\Delta V \left( Q_I^{(0)} + Q_I^{(1)}
  + Q_I^{(2)} \right)} \leq 1, \label{eq:eta_em}
\end{equation}
in consistence with Eqs.~(\ref{eq:energycons}) and (\ref{eq:2ndlaw}) for a truncation in the frequency expansion up to first order in the CIF, which implies a second-order term in the currents~\cite{juergens2013,bustos2019}.
In the opposite case where $\mathcal{W}_\tf{ext} < 0$, now the input and output energies swap roles, so the efficiency of this \textit{electrical pump} is
\begin{equation}
  \eta_\tf{ep} = - \frac{\Delta V \left(Q_I^{(0)}+Q_I^{(1)}+Q_I^{(2)}\right)}{\mathcal{W}_\tf{ext}} \leq 1.
  \label{eq:eta_ep}
\end{equation}
Such a quantity, however, is only well defined in the case where the total amount of transported charge per period $Q_I$ is opposed to that given by the natural direction of the bias current. In the used sign convention for the charge currents this means that $Q_I \Delta V > 0$.

A similar analysis can be done in the case where one replaces the bias voltage by a temperature gradient between the contacts, such that the device can operate either as a heat engine or a refrigerator.
By establishing different temperatures in the leads, defined as $T_\tf{hot} = T+\Delta T/2$ and $T_\tf{cold} = T-\Delta T/2$, a heat current flows through the DQD system which, in turn, may activate its mechanical component. In this scenario where $\mathcal{W}_\tf{ext} > 0$, the device is driven by the heat current coming from the hot lead, $-Q_{J_\tf{hot}}$. This means that the device operates as a \textit{heat engine} with efficiency
\begin{equation}
  \eta_\tf{he} = -\frac{\mathcal{W}_F^{(0)} + \mathcal{W}_F^{(1)}}
  {Q_{J_\tf{hot}}^{(0)}+Q_{J_\tf{hot}}^{(1)}+Q_{J_\tf{hot}}^{(2)}}
  \leq \frac{\Delta T}{T_\tf{hot}},
\label{eq:bound_hm} 
\end{equation}
where $Q_{J_\tf{hot}}$ is defined as the time integral of $J_\tf{hot}$ over a period given by $\Omega$, while $J_\tf{hot}$ is taken up to second order in this quantity. On the other hand, when $\mathcal{W}_\tf{ext} < 0$ the heat current flows against the temperature gradient. Assuming that the total amount of transported heat to the cold reservoir is negative, $Q_{J_\tf{cold}} < 0$, we can define the efficiency (or coefficient of performance) of this \textit{heat pump} or \textit{refrigerator} by the expression:
\begin{equation}
  \eta_\tf{hp} = \frac{Q_{J_\tf{cold}}^{(0)}+Q_{J_\tf{cold}}^{(1)}+Q_{J_\tf{cold}}^{(2)}}{\mathcal{W}_\tf{ext}} \leq \frac{T_\tf{cold}}{\Delta T},
\label{eq:bound_hp}
\end{equation}
where again the heat current $J_\tf{cold}$ is taken up to second order in $\Omega$~\cite{bustos2019}.

Finally, it is convenient to define normalized efficiencies with respect to the maximum theoretical value, given by Eqs.~(\ref{eq:bound_hm}) and (\ref{eq:bound_hp}), i.e.,
\begin{equation}
\tilde{\eta}_\tf{he} = \frac{T_\tf{hot}}{\Delta T}\eta_\tf{he}, \quad \tf{and} \quad \tilde{\eta}_\tf{hp} = \frac{\Delta T}{T_\tf{cold}}\eta_\tf{hp} .
\end{equation}

\subsection{Decoherence model}
\label{subsec:decoh}

One of the key questions motivating this work is whether quantum coherence plays a role in the operation of QD-based nanodevices such as adiabatic quantum motors and pumps. In this regard, studying the effect of decoherence on the machines' performance is crucial.

Calculating decoherent relaxation times from a microscopic theory would require identifying the dephasing mechanisms, which is beyond the scope of this work. Instead, we choose a phenomenological approach~\cite{engel2002,cota2005} that consists of inserting the relaxation times directly into the master equations.
In our case, this implies adding to the kernel $\bm{W}^{\tf{eff}}$ a decoherence rate $\Gamma_\phi$.~\footnote{It is straightforward to include in the model of decoherence also a thermalization mechanism. However, for the particular example used here, this effect does not have any impact on the system as we are in the limit of degenerate quantum levels.}
The inclusion of $\Gamma_\phi$ is only done in the diagonal elements of the $\tf{nn}$ block of $\bm{W}^{\tf{eff}}$, i.e., $[\bm{W}_\tf{nn}^{\tf{eff}}]_{ii} \rightarrow [\bm{W}_\tf{nn}^{\tf{eff}}]_{ii} - \Gamma_\phi$.
This phenomenological rate describes any decoherent process that may occur in the quantum dots, present even in the absence of a coupling to the leads. This type of decoherence destroys the information about the relative phase in a superposition of states $\alpha$ and $\beta$ ($p_{\alpha,\beta}$) without changing the populations of the states ($p_{\alpha,\alpha}$ and $p_{\beta,\beta}$).
Without a coupling to the reservoirs, this formally leads to a decay of the off-diagonal matrix element $p_{\alpha,\beta}(t)$.
In our case, however, there is also a replenishing mechanism given by the fact that when electrons enter into the system, they do it in a superposition state. Therefore, it is expected that coherences $p_{\alpha,\beta}$ reach a $\Gamma_\phi$-dependent steady state at long times.
In the following sections, we will take $\Gamma_{\phi}$ as an ``external knob'' that can be used to test the effect of decoherence on the machines' performance.

\section{DQD in the weak interdot coupling regime}
\label{sec:dqd}

In this section we will apply the formalism and assumptions described previously to the particular
example of a double dot weakly coupled to two external leads and capacitively coupled to a rotor.

\subsection{Hamiltonian and physical model}
\label{subsec:physical}

The local system we are about to study is a DQD device composed of two single-level spin-degenerate
quantum dots coupled to each other, together with a rotative mechanical piece placed in their
proximity and capacitively coupled to them.
This will be the only type of coupling considered between the dots and the rotor; i.e., tunneling events between these subsystems are 
not taken into account.
At the same time, the whole device is weakly coupled to
source (S) and drain (D) leads, as depicted in Fig.~\ref{fig:scheme}. By weak 
coupling we mean that the broadening due to tunneling events is 
much smaller than the temperature broadening, i.e., $\Gamma \ll k_\tf{B}T$. Notice that, depending 
on the choice of the tunnel rates $\Gamma_{r\ell}$, it is possible to configure the double quantum 
dot arrangement either in series or in parallel [see Figs.~\ref{fig:scheme}(c) and \ref{fig:scheme}(d)]. The 
asymmetry between source and drain rates is quantified by the factor
\begin{equation}
\lambda = (\Gamma_\tf{S}-\Gamma_\tf{D})/\Gamma, 
\label{eq_lambda}
\end{equation}
where $\Gamma_r = \Gamma_{r1} + \Gamma_{r2}$. In addition, for a specific lead $r=\{ \tf{S}, \tf{D} \}$, we define 
the lead-dot asymmetry factor as
\begin{equation}
\lambda_r = (\Gamma_{r1} - \Gamma_{r2}) / \Gamma_r. 
\label{eq_lambda_r}
\end{equation}
These factors will be useful later on for setting different system configurations and for the search 
of a suitable working point (see Secs.~\ref{subsec:coherences} and \ref{sec:example}). The local system 
is represented by the electronic Hamiltonian
\begin{equation}
\begin{aligned}
  \hat{H}_\tf{el} = & \sum_\ell E_{\ell} \hat{n}_{\ell} + U \hat{n}_1 \hat{n}_2 + \frac{U'}{2} \sum_\ell \hat{n}_{\ell} (\hat{n}_{\ell} - 1) \\
					& -\frac{t_c}{2} \sum_\sigma (\hat{d}_{1\sigma}^\dag \hat{d}_{2\sigma} + \tf{H.c.}), 
\end{aligned}
\label{eq:helddot}
\end{equation}
where $\hat{n}_\ell$ is the $\ell$-dot particle number operator, defined as
$\hat{n}_{\ell} = \sum_\sigma \hat{d}_{\ell \sigma}^\dag \hat{d}_{\ell \sigma}$, while
$E_{\ell} = E_{\ell} (\bm{X})$ represents the on-site energy of each dot $\ell = \{1, 2\}$,
which is locally tuned by its coupling to the mechanical part of the system. $t_c$
denotes the interdot coupling amplitude while $U$ and $U'$ represent the inter- and intradot Coulomb
interactions, respectively.
For the sake of simplicity, we will take in the following these parameters to be much larger than all other energy scales in the system ($U, U' \rightarrow \infty$), such that the double-dot device can only be singly occupied or empty.
Due to these assumptions, the only states relevant for our system are $\ket{0}$ and $\ket{\ell\sigma}$, where the former means that both quantum dots are empty and the latter that there is one electron with spin $\sigma$ in the dot $\ell$.

Applying a bias voltage and/or a temperature gradient between the leads will cause charge and heat 
to flow through the dots.
If the mechanical piece is coupled to the DQD then an energy exchange between these subsystems is possible.
As Figs.~\ref{fig:scheme}(a) and \ref{fig:scheme}(b) suggest, the cyclic motion of the rotor (which can be thought of as an electrical dipole) modifies the quantum dots energy levels, similarly to the action of externally controlled gate voltages.
In agreement with Eqs.~(\ref{eq:energycons})-(\ref{eq:onsager_heat}), once a bias voltage or a temperature gradient is applied, the current flowing through the dots releases part of its energy to the mechanical subsystem making it rotate.
The opposite scenario can be achieved by applying an external force into the mechanical system, such that its motion produces a finite current through the electronic device.
In Appendix~\ref{app:trajectory} we discuss in more detail the example shown in Fig.~\ref{fig:scheme} and how it might be possible to control the coupling between the quantum dots and the rotor.
Another possibility could be a dipolar molecule in proximity to the quantum dots such that there exists a capacitive coupling between the subsystems.
In any case, for the purpose of the present work, what really matters is not the specific details of the mechanical system, but its effects on the electronic Hamiltonian.
What enables this energy conversion is the dependence of the energy levels of the dots on the position of the mechanical rotor, which in this case can be characterized by an angle $\theta$.
This $\theta$ dependence is related to physical characteristics such as the rotor's length and
its position with respect to the DQD, and the coupling strength between the rotor and the dots. A strict derivation of this angular dependence requires an accurate knowledge of the
rotor's details, which can yield complex parametrizations for the dots' on-site energies.
As the aim of the work is to unveil the role of coherences on CIFs and not to focus on specific details of a particular device, we assume a simple $\theta$ dependence for the dots' energies, given by
\begin{align}
  E_1 (\theta) &= \bar{E}_1 + \delta_E \cos(\theta) + \delta_\epsilon \sin(\theta), \nonumber\\
  E_2 (\theta) &= \bar{E}_2 + \delta_E \cos(\theta) - \delta_\epsilon \sin(\theta), 
\end{align}
where $\delta_E$ and $\delta_\epsilon$ describe the electromechanical coupling.
According to the model shown in Fig.~\ref{fig:scheme} and discussed in Appendix~\ref{app:trajectory}, they are related to the capacitances acting on the DQD.
In the energy space, these equations describe an elliptic trajectory  of radius $\delta_E$ and $\delta_\epsilon$ around the \textit{working point} $(\bar{E}_1, \bar{E}_2)$.
This trajectory is convenient given the typical shape of the curvatures for the configuration of interest of the DQD; see 
Fig.~\ref{fig:parallelBI}(a) for example.
We assume that these energies $\bar{E}_{\ell}$ can be externally tuned (for example, by external gate voltages) so that the working point can be chosen favorably. If we are thinking in the performance of motors or pumps, then this convenience lies in the fact that, to get useful work or pumped charge/heat,
we need to find some region in the parameter space where their associated curvatures are
nonzero (cf. Sec. \ref{subsec:eta}). Obviously, the above parametric approach also applies to the energy difference $\epsilon = E_1 - E_2$ and the mean level energy $E = (E_1+E_2)/2$, such that these can also be treated as tunable parameters through the following equations:
\begin{equation}
  E (\theta) = \bar{E} + \delta_E \cos(\theta), \; \tf{and} \; \epsilon (\theta) = \bar{\epsilon} + 2 \delta_\epsilon \sin (\theta).
\end{equation}
Importantly, at the level of approximation used in this work, the energy difference between the dots needs to be taken perturbatively, 
i.e., $\epsilon \sim \Gamma$. As we shall see next, the regions in which the curvature associated with the CIF is nonzero lies below 
this constraint, such that we can safely define a trajectory enclosing the relevant region of $\mathcal{B}^F$ with $\delta_\epsilon$ 
on the order of $\Gamma$.

\subsection{Regime of parameters}
\label{subsec:regime}

With the purpose of studying the potential role of quantum coherences on these devices, we will now focus on the weak interdot coupling regime where $t_c \sim \Gamma$.
Adiabatic quantum motors, heat engines, and charge/heat pumps in the strong coupling regime ($t_c \gg \Gamma$) have already been addressed~\cite{riwar2010,juergens2013,calvo2017,bustos2019}.
There, it was shown that coherences have no important contributions to any of the quantities of interest (e.g., charge and heat currents, CIFs, etc.) and can therefore be disregarded to lowest order in $\Gamma$.
On the contrary, in the weak-coupling regime, the role of the coherent superposition among the DQD states becomes crucial for the operation of electron pumps. This was studied in Ref.~[\onlinecite{riwar2010}]. Due to the connection between adiabatic quantum motors and pumps [cf. Eqs.~(\ref{eq:energycons})--(\ref{eq:onsager_heat})], it is expected that coherent effects are also relevant for the performance of quantum motors and heat engines in the weak-coupling regime.

Whether or not a system is in the weak or in the strong coupling regimes depends on the comparison between $\Gamma$ and the energy difference between the eigenstates of the system. When this difference is much bigger than $\Gamma$, coherent effects can be disregarded, at least to the lowest order in $\Gamma$~\cite{wunsch2005,leijnse2008}.
Here we are in the opposite case, which occurs when both $\epsilon$ and $t_c$ are of the order of $\Gamma$.
The assumption implies that single-electron states are almost degenerate and guarantees the coherences' survival~\cite{riwar2010}, laying the ground for the study of their potential effect on autonomous quantum machines like the one studied here.

With respect to the kernel $\boldsymbol{W}^\mathrm{eff}$, see Eqs.~(\ref{eq:kernel}), (\ref{eq:W^eff}) and Appendix~\ref{app:kernel}, all kernel blocks depend on the mean level energy $E$.
However, the $\epsilon$ dependence only enters in the $\bm{W}_\tf{nn}^\tf{eff}$ block, which contains local information of the system through the Liouvillian $\bm{L}$.
As discussed before in Sec.~\ref{subsec:eta}, the geometrical nature of the first-order pumped charge and heat, and the zeroth-order work of the CIFs, implies that a two-parameter dependence is necessary for these quantities to be nonzero.
In our system, this condition can only be fulfilled if there is a coupling between the diagonal and nondiagonal blocks of $\bm{W}^\tf{eff}$. 
The coupling of the kernels' blocks inevitably leads, in turn, to the coupling of occupations and coherences of the reduced density matrix.
Therefore, we can state that occupations and coherences need to be coupled to have finite pumping/work in DQD-based nanodevices within the weak-coupling regime.

Note that since we work in a regime where $t_c \sim \epsilon \sim \Gamma$, taking up to first-order terms in these parameters implies that the exact bonding/antibonding basis reduces to the local energy basis. Specifically, the system's eigenstates need to be taken into account in zeroth order in these parameters and equal those of the fully decoupled single dots $\ket{1\sigma}$ and $\ket{2\sigma}$; see Refs.~[\onlinecite{wunsch2005}] and [\onlinecite{riwar2010}] for more details.
The vector $\bm{p}$ then adopts the form $\bm{p}= \left( p_{0}, p_{1\uparrow}, p_{1\downarrow}, p_{2\uparrow}, p_{2\downarrow}, p_{1\uparrow,2
\uparrow}, p_{1 \downarrow,2\downarrow}, p_{2\uparrow,1\uparrow}, p_{2\downarrow,1\downarrow} \right)^\tf{T}$, where we use $p_\alpha = p_{\alpha,\alpha}$. Its components represent the
occupation probabilities for the device to be either in the empty state $\ket{0}$ ($p_0$) or in
the singly occupied state $\ket{\ell \sigma}$ ($p_{\ell \sigma}$), and coherent
superpositions between single-particle eigenstates $\ket{\ell \sigma}$ and $\ket{\ell'\sigma}$ ($p_{\ell \sigma,\ell' \sigma}$).
As we shall see later on, when the DQD is connected in series, such a coupling will be provided by $t_c$. However, in the parallel configuration we will see that even in the absence of $t_c$, the coupling between $\bm{p}_\tf{d}$ and $\bm{p}_\tf{n}$ still holds. This is due to the fact that, under this condition, electrons coming from the leads enter into the DQD in a coherent superposition of states.

\subsection{Role of coherences and the decoupled parallel configuration}
\label{subsec:coherences}

\begin{figure}[th!]
  \centering
  \includegraphics[width=1.\columnwidth]{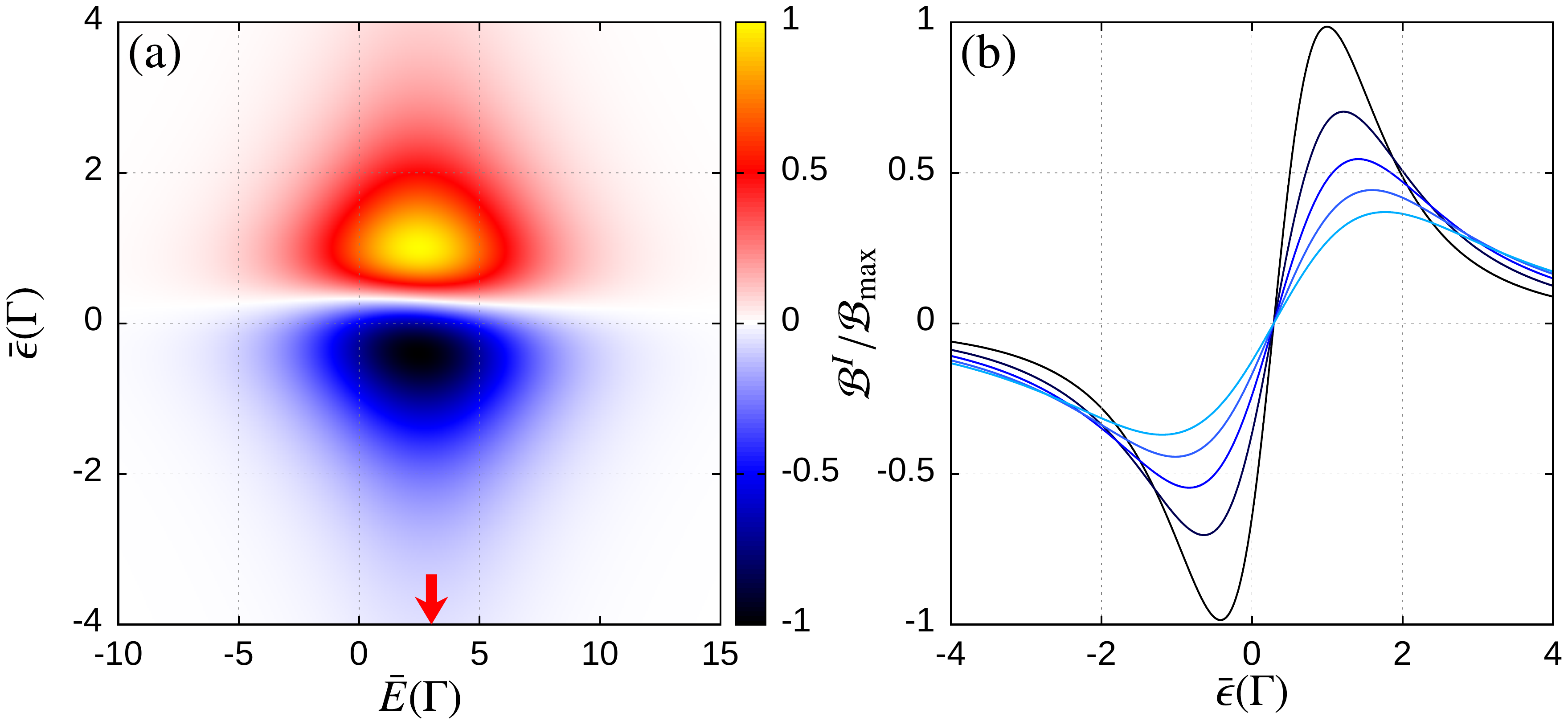}
  \caption{(a) Map of the charge current curvature $\mathcal{B}^I$ as a function of $\bar{E}$ and $\bar{\epsilon}$ for a DQD in series and in the 
  absence of bias voltages and temperature gradients. The shown curvature is normalized to its maximum absolute value within the shown map, 
  $\mathcal{B}_\tf{max} \sim 0.06 \, e/(k_\tf{B} T)^2$. (b) A cut of the curvature for $\bar{E} = 3 \Gamma$ [see red arrow in (a)] and for several   
  decoherence rates: $\Gamma_\phi = 0, 0.25, 0.5, 0.75$, and 1, in units of $\Gamma$. The darkest curve corresponds to the case where there is no   
  decoherence ($\Gamma_\phi = 0$) while the lightest one denotes the case of highest decoherence rate ($\Gamma_\phi = \Gamma$).
  The rest of the curves are for intermediate values of $\Gamma_\phi$.
  The other parameters used are $\Gamma = t_c = 0.5 \, k_\tf{B} T$, $\lambda = 0.5$, $\lambda_\tf{S} = 1.0$, and $\lambda_\tf{D} = - 1.0$.}
\label{fig:seriesBI}
\end{figure}
A serially coupled DQD in the weak interdot coupling regime was considered in Ref.~[\onlinecite{riwar2010}]. 
There it was shown that the system is capable of pumping charge without an applied bias voltage if the 
DQD is asymmetrically coupled to the leads ($\lambda \neq 0$).
This can be seen in Fig.~\ref{fig:seriesBI}(a) where we show a map of the charge current curvature $\mathcal{B}^I$ as a function of $\bar{E}$ and $\bar{\epsilon}$.
This quantity allows one to determine those regions in the space of parameters over which a closed trajectory can be traced for the production of a net pumped charge current after one modulation cycle~\cite{calvo2012}.
In the figure, we observe a two-lobe pattern with opposite signs. The shift of sign of the current curvature is due to a renormalization of energy levels attributed to the Coulomb interaction~\cite{wunsch2005,riwar2010}.
As discussed in Sec.~\ref{subsec:eta}, moving the parameters $E$ and $\epsilon$ such that their trajectory encircles any region of Fig.~\ref{fig:seriesBI}(a), without a change of sign, ensures a finite pumped charge. This alone proves that the system can be used as an adiabatic quantum pump.

The configuration in series of the DQD, see Figs.~\ref{fig:scheme}(a) and \ref{fig:scheme}(c), allows one to access different operational modes of the device, i.e., adiabatic quantum motors, adiabatic quantum pumps, etc.
However, in this case, quantum coherences come entirely from the coupling between the quantum dots.
This is clear also when one analyses the structure of the effective kernel $\bm{W}^\tf{eff}$. There, for the series configuration, the $\tf{nd}$ and $\tf{dn}$ blocks of the effective kernel are only given by the local Hamiltonian, such that:
\begin{equation}
\bm{W}^\tf{eff}_\tf{dn/nd} = -\tf{i}\bm{L}_\tf{dn/nd}.
\end{equation}
These blocks are responsible for the arising of coherences and for the coupling with the $\bm{W}^\tf{eff}_\tf{nn}$ block (which ultimately leads to finite pumping). In this configuration, all matrix elements in $\bm{W}^\tf{eff}_\tf{dn/nd}$ are proportional to $t_c$; i.e., there are no contributions from the evolution kernel $\bm{W}$. Thus, e.g., taking $t_c = 0$ not only destroys any coherent superposition but also trivially cuts charge/heat transport through the local system.
In this sense, the parallel configuration of the DQD, see Fig.~\ref{fig:scheme}(b), offers a richer example to study the role of coherences.
There, coherences do not solely come from the interdot coupling but also from the particles entering simultaneously to both dots. This comes from the fact that novel tunnel processes are enabled, since the matrix elements of $\bm{W}_\tf{nd/dn} \propto t_{r,1} t_{r,2}^{*}$ become nonzero (see Appendix~\ref{app:kernel}).

\begin{figure}[th]
  \centering
  \includegraphics[width=1.\columnwidth]{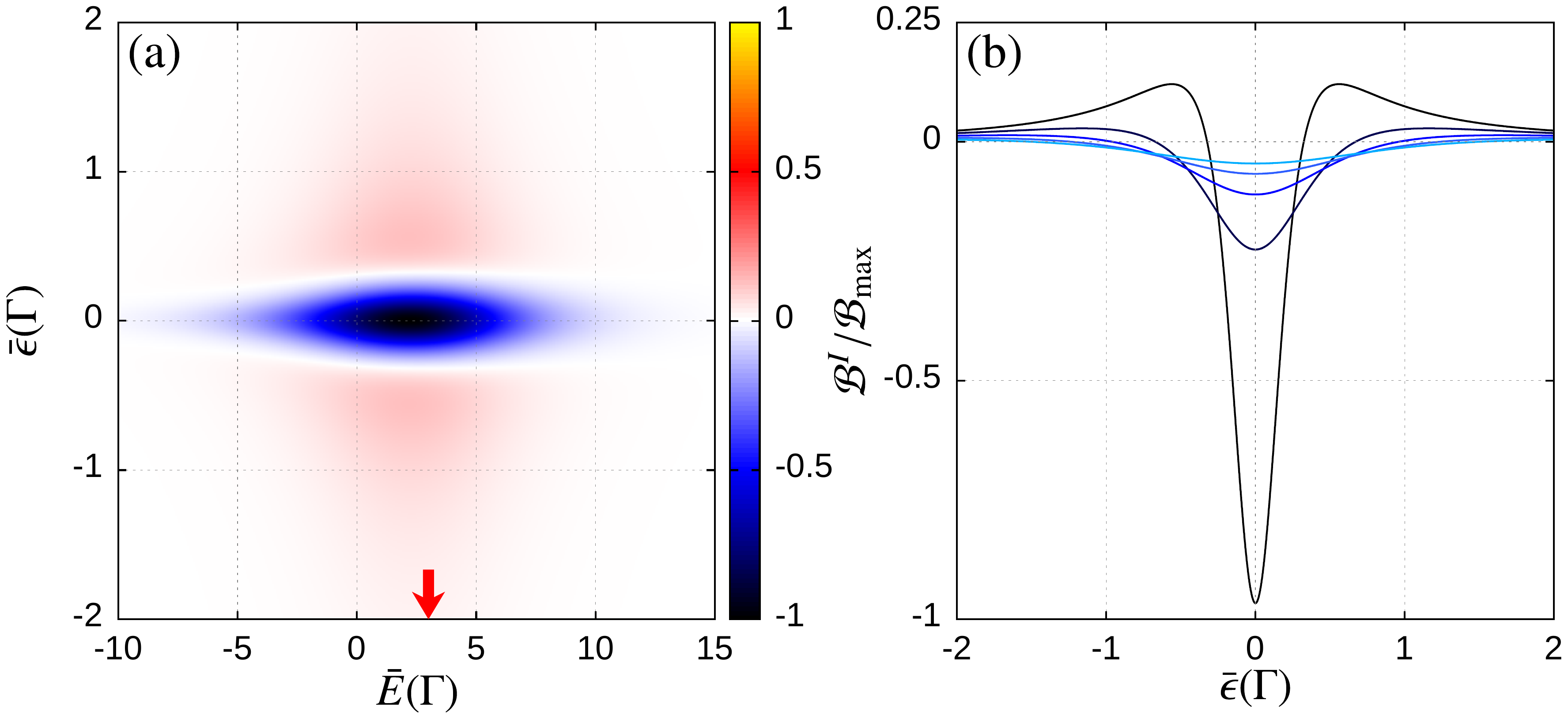}
  \caption{(a) Map of the charge current curvature $\mathcal{B}^I$ as a function of $\bar{E}$ and $\bar{\epsilon}$
  for a DQD in parallel and in the absence of bias voltages and temperature gradients. The shown curvature is normalized to its maximum absolute value 
  $\mathcal{B}_\tf{max} \sim 0.5 \, e/(k_\tf{B} T)^2$ whithin the shown map.
  (b) A cut of the curvature for $\bar{E} = 3 \Gamma$  [see red arrow in (a)] and for several decoherence rates: $\Gamma_\phi = 0, 0.25, 0.5, 0.75$, and 1, in units of 
  $\Gamma$. The darkest curve corresponds to the case where there is no decoherence ($\Gamma_\phi = 0$) while the lightest one denotes the 
  case of highest decoherence rate ($\Gamma_{\phi} = \Gamma$). The rest of the curves are for intermediate values of
  $\Gamma_\phi$. The used parameters are $\Gamma = 0.5 \, k_\tf{B}T$, $t_c = 0$, $\lambda = 0$, $\lambda_\tf{S} = 0.5$, 
  $\lambda_\tf{D} = - 0.5$.}
\label{fig:parallelBI}
\end{figure}

In contrast to the configuration in series, in the parallel scenario it is possible to pump charge or heat even if the dots are decoupled ($t_c = 0$); see Fig.~\ref{fig:parallelBI}(a). This particular case does not have a classical analog as classical particles entering one dot do not have a way of getting information from the other one. Then, a ``classical'' DQD should behave as two independent single-parameter systems and, because of that, the total pumped charge, e.g., should be zero. Therefore, the quantum nature of electrons, which allows for a coherent superposition of the wave functions, is what is ultimately responsible for the pumping of charge and heat, and the production of finite work from the CIFs. One can interpret that quantum coherences are what allows a particle to get information from both dots making the pumping depend on two parameters, $E_1$ and $E_2$ (or $E$ and $\epsilon$). In this sense, devices based on this configuration can be considered as “true” quantum machines.

An important aspect of the coherent superposition of the dots' states should be mention here. To some extent, the interaction between the mechanical and electronic components of the system takes place through a measurement on the level occupations of the DQD. Such a measurement by the classical part of the system would destroy any superposition state existing in the DQD. Since we argue that in this system the work extraction (or pumped charge/heat)  relies on the coherent superpositions, and these are destroyed in order to know the dots' occupations, it might seem like there is a contradiction here. However, one should notice at this point that the occupations \text{already} contain information on the superpositions. The occupations are not separated from the coherences, as would be the case in the sequential tunneling regime. Instead, they can be written as the sum of two terms (see Appendix~\ref{app:kernel}). One of them corresponds to the \textit{incoherent} contribution to the occupations: these are the ones that would be obtained in a sequential tunneling regime. The second term corresponds to the coherent contribution which comes from the superposition of the DQD states. It is precisely this term that is related to the work extraction and adiabatic pumping. To give a rather physical interpretation, one can imagine that the electron enters into the DQD system in a superposition state. Of course, when the \textit{slow} mechanical system interacts with the electronic component, this superposition gets destroyed, but the probabilities for the final electron states $\ket{\ell \sigma}$ are still defined by the above two contributions. In summary, since the classical component is assumed to be much slower than the typical electron timescales, the resulting work done by the CIF (or pumped charge/heat) comes from the average of the statistical ensemble of the many electrons passing through the device, an average that in turn contains information about the coherent superposition of the dots.

Before analyzing the effect of decoherence on the series and the decoupled parallel configurations of the DQDs, we want to analyze an interesting case of the system's parameters that leads to zero pumping. As we pointed out before, the coupling between the $E$-dependent block of the effective kernel ($\boldsymbol{W}^\mathrm{eff}_\mathrm{dd}$) and the $\epsilon$-dependent one ($\boldsymbol{W}^\mathrm{eff}_\mathrm{nn}$) is what provides the two parameters needed for finite pumping. However, even in the presence of such a coupling, we found that setting $\lambda_\tf{S} = \lambda_\tf{D}$ leads to zero pumping, in the absence of a bias voltage or a temperature gradient, independently of the choice of $\lambda$.
This occurs despite the fact that there is no obvious inversion symmetry [see Eqs.~(\ref{eq_lambda}) and (\ref{eq_lambda_r})], even when parameters $E$ and $\epsilon$ enclose a finite area in the parameters' space.
However, this particular case can be understood once one realizes that the pumping currents become proportional to each other: $I_r^{(1)} = (\Gamma_r/\Gamma_{r'}) I_{r'}^{(1)}$, where $r, r' = \{\tf{S},\tf{D}\}$; see Appendix~\ref{app:symmetry}.
The proportionality between both currents implies that each one can be written as the total time derivative of the average zeroth-order occupation number $\braket{\hat{n}}$ and, thus, they integrate to zero for a whole cycle; see Appendix~\ref{app:symmetry}.

As a way to test the role of coherences on DQD-based quantum machines in the series configuration, we show in Fig.~\ref{fig:seriesBI}(b) the current curvature for a fixed mean energy, $\bar{E} = 3 \Gamma$, and different decoherence rates $\Gamma_\phi$. There it can be seen that $\Gamma_\phi$ produces an amplitude decay and a widening of the curvature peaks which can be attributed to a gradual attenuation in the coherent coupling between the quantum dots. Interestingly, for intermediate values of $\Gamma_\phi$ there are some regions in parameter space where decoherence increases the magnitude of the curvature due to its broadening effect; see for example $\bar{\epsilon} \sim 3 \Gamma$.

In Fig.~\ref{fig:parallelBI}(a) we show the current curvature $\mathcal{B}^I$ but for the decoupled parallel configuration. Now, we observe a three-lobe pattern but dominated by a single sign (here the absolute value of the central peak is much greater than that corresponding to the side peaks).
The role of $\Gamma_\phi$ on the curvature, see Fig.~\ref{fig:parallelBI}(b), is the same as in the configuration in series, but the fact that $t_c = 0$ in this case allows us to give a more direct interpretation of its effect for sufficiently large values. In this limiting situation, the two dots become effectively decoupled since the characteristic survival time of the superposition now goes like $1/\Gamma_\phi$. Then, each dot is unaware of the other dot's existence as all phase information gets lost much faster than the typical time spent by the electron in the DQD system.
As mentioned, this leads to a monoparametric scheme (with $E$ the only parameter being modulated), such that $\mathcal{B}^I \rightarrow 0$, and therefore no working device can be created.

\begin{figure*}[ht!]
  \centering
  \includegraphics[width=0.8\textwidth]{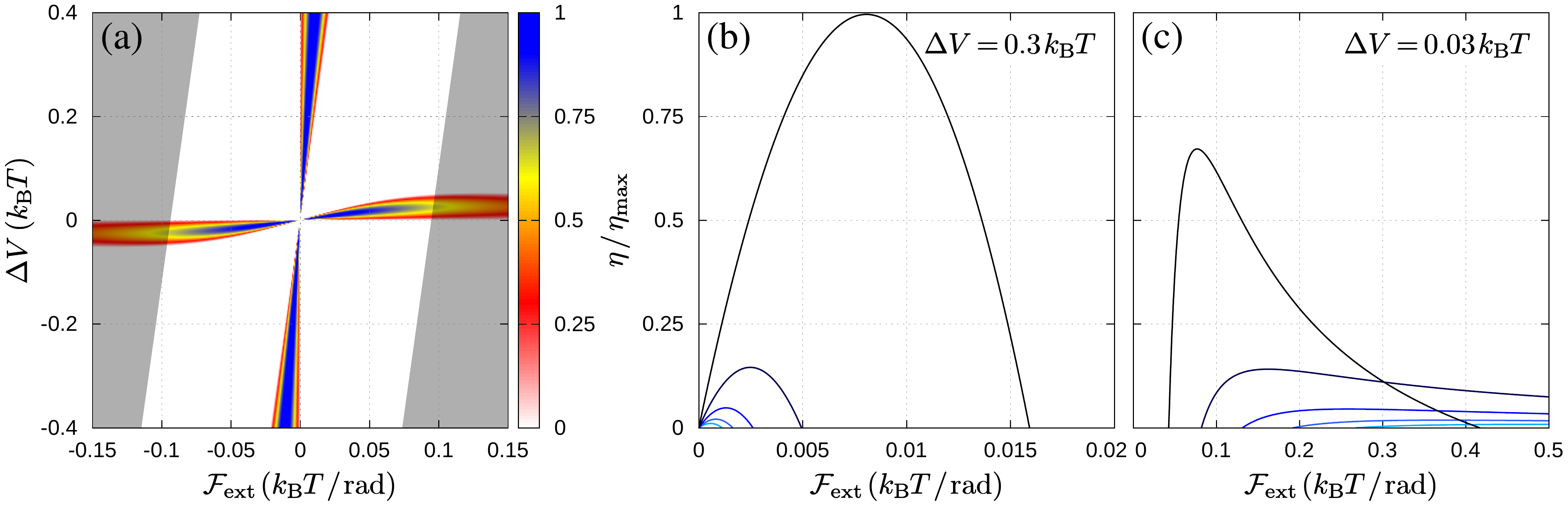}
  \caption{(a) Map of the efficiencies $\eta_\tf{em}$ and $\eta_\tf{ep}$ as functions of the external torque $\mathcal{F}_\tf{ext}$ and the bias
  voltage $\Delta V$. The efficiencies have been normalized with respect to the maximum value $\eta_\tf{max} = 1.05 \times 10^{-2}$ achieved in 	
  the shown region. Shaded areas denote the regions where the adiabaticity condition is not fulfilled (see main text). (b) Electric motor efficiencies 	 as a function of $\mathcal{F}_\tf{ext}$ for $\Delta V = 0.3 \, k_\tf{B}T$ and different decoherence rates in units of $\Gamma$: $\Gamma_\phi =$ 0
  (black), 0.25, 0.5, 0.75, and 1 (cyan). (c) Electric pump efficiencies as a function of $\mathcal{F}_\tf{ext}$ for $\Delta V = 0.03 \, k_\tf{B}T$  
  and the same values of $\Gamma_\phi$ as in (b). The other used parameters are $\Gamma = 0.5 \, k_\tf{B} T$, $t_c = 0$, $\lambda = 0$, 
  $\lambda_\tf{S} = 0.5$, and $\lambda_\tf{D} = -0.5$. The chosen trajectory is given by $\bar{E} = 1.1 \, k_\tf{B} T$, $\bar{\epsilon} = 0$, 
  $\delta_E = 5 \, k_\tf{B}T$, and $\delta_\epsilon = 0.15 \, k_\tf{B} T$.}
  \label{fig:etavb}
\end{figure*}
\begin{figure}[!h]
  \centering
  \includegraphics[width=1.0\columnwidth]{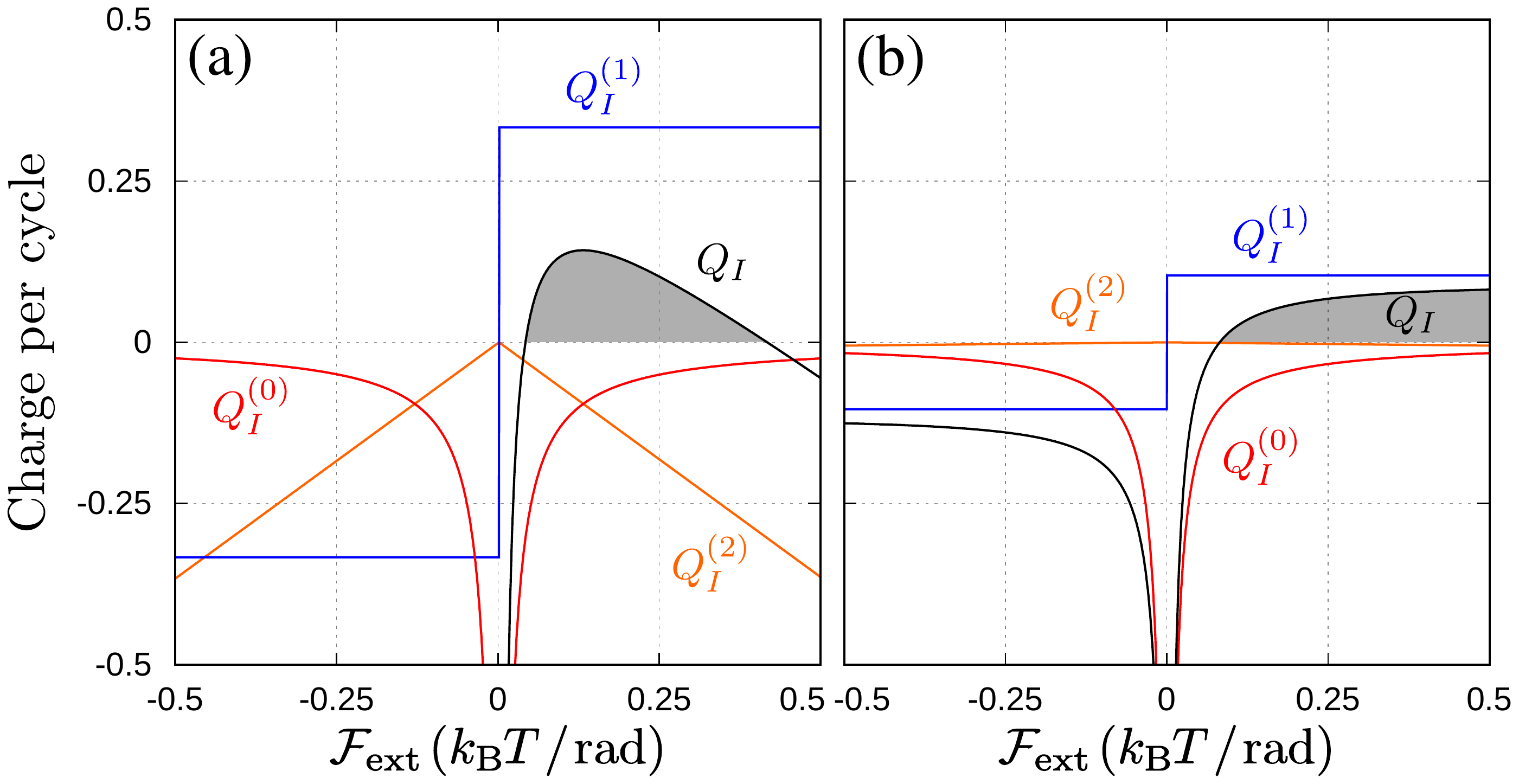}
  \caption{Different order contributions to the transported charge as a function of the external torque $\mathcal{F}_\tf{ext}$. The sum of all these 
  contributions, denoted by $Q_I$, is shown in solid black. The gray area indicates the region where the device is capable of pumping charge and 
  therefore becomes operational. We considered the same parameters as in Fig.~\ref{fig:etavb}(c) with $\Gamma_\phi = 0$ (a) and 
  $\Gamma_\phi = 0.25 \Gamma$ (b).}
\label{fig:charge}
\end{figure}

\section{Quantum machines based on the decoupled parallel configuration}
\label{sec:example}

In the previous section, we discussed the role of coherence in charge pumping, but the studied device admits other operational regimes. Here, we study the effects of applying an external driving force together with a bias voltage or a temperature gradient to the decoupled parallel configuration. This takes the system into different operational regimes, namely, electric motor, charge pump, heat engine, or heat pump. We start by describing the effect of a bias voltage and an external force.

The shown charge current curvatures in Figs.~\ref{fig:seriesBI} and \ref{fig:parallelBI} are clear indications that the device may operate as a charge pump in the situation where there is no applied bias voltage. Although not shown, the force curvature $\mathcal{B}^F$ displays a similar pattern thus implying that the device could also work as an electrical motor for the chosen parameters. However, to understand in more detail the device's operational behavior, we need to take into account the effect of the external force $\bm{F}_\tf{ext}$. This force will finally determine whether the system is operational or not, together with its subsequent working mode, i.e., motor or pump.
For simplicity we assume that the external force is constant and points along the tangential direction $\hat{\bm{\theta}}$, i.e., $\bm{F}_\tf{ext} = F_\tf{ext} \hat{\bm{\theta}}$, so its associated torque $\mathcal{F}_\tf{ext}$ is constant along the whole trajectory and we can take $\mathcal{W}_\tf{ext} = 2\pi s \mathcal{F}_\tf{ext}$. Let us recall that $s = \pm 1$ gives the dynamically determined direction of rotation, since $\dot \theta = s |\dot \theta|$.

One of the interesting aspects of the system analyzed here is that one can control its operational mode externally, switching between a pump and a motor by moving the bias voltage or the externally applied torque. 
For the case of the motor, the system will act as such when the external work is positive, which means that $s$ and $\mathcal{F}_\tf{ext}$ need to have the same sign. However, although $\mathcal{F}_\tf{ext}$ can be controlled, this \textit{a priori} does not determine $s$. As Eq.~(\ref{eq:dottheta}) states, the direction of rotation also depends on the CIF's coefficient. For the case of the charge pump, on the other hand, an additional condition to $\mathcal{W}_\tf{ext}<0$ is required: The system will act as a proper pump when the first-order current overcomes the leakage currents, typically given by the zeroth- and second-order contributions. This implies $|Q_I^{(1)}| > |Q_I^{(0)} + Q_I^{(2)}|$. Recalling that $Q_I^{(k)} \propto \Omega^{k-1}$ and Eq.~(\ref{eq:dottheta}), it is clear that the external torque ultimately controls the compensation between the leakage and pump currents. One can also take the voltage as the pump knob. Here, one is mainly controlling the zeroth-order current, proportional to $\Delta V$, at low bias.

\begin{figure*}[th]
  \centering
  \includegraphics[width=0.8\textwidth]{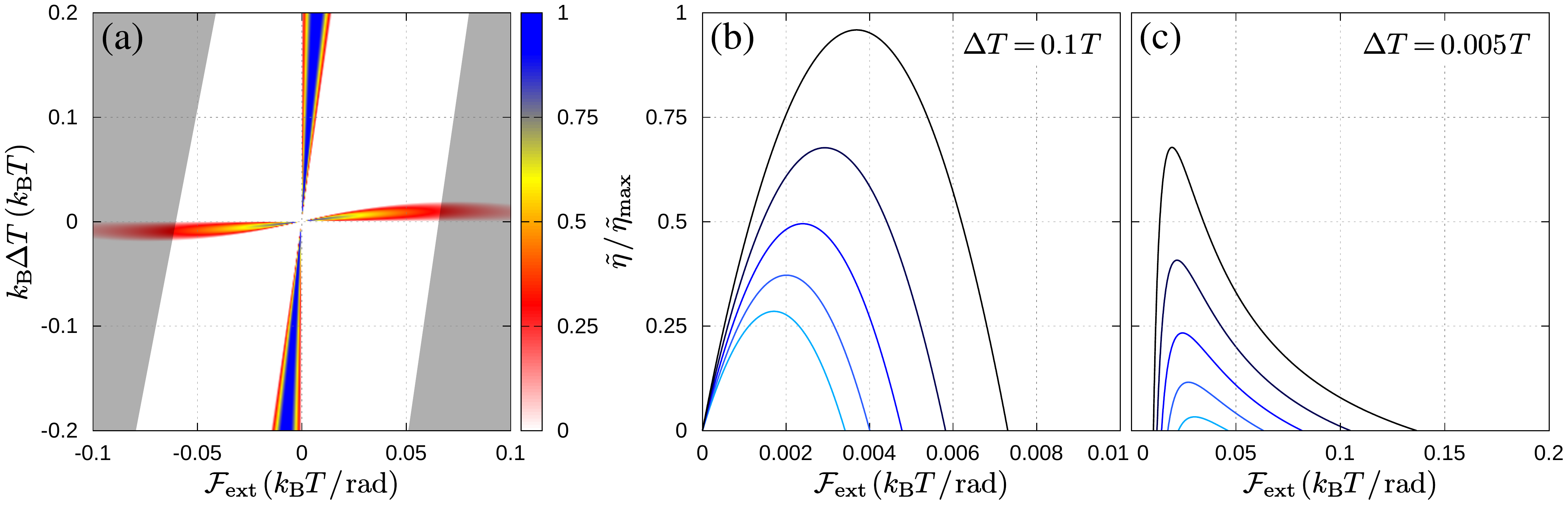}
  \caption{(a) Map of the \textit{normalized} efficiencies $\tilde{\eta}_\tf{he}$ and $\tilde{\eta}_\tf{hp}$ as a function of 
  $\mathcal{F}_\tf{ext}$ and $\Delta T$. These functions have been divided with respect to the maximum value 
  $\tilde{\eta}_\tf{max} = 1.04 \times 10^{-2}$ achieved in the shown map. As in Fig.~\ref{fig:etavb}, the shaded areas denote the regions where the 
  adiabaticity condition is not satisfied. (b) Plots of $\tilde{\eta}_\tf{he}$ vs $\mathcal{F}_\tf{ext}$ for $\Delta T = 0.1 \, T$ and for different 
  decoherence rates (in units of $\Gamma$): $\Gamma_\phi =$ 0 (black), 0.05, 0.1, 0.15, and 0.2 (cyan). (c) Plots of $\tilde{\eta}_\tf{hp}$ vs 
  $\mathcal{F}_\tf{ext}$ for $\Delta T = 0.005 \, T$ and for the same values of $\Gamma_\phi$ as in (b). The other used parameters are: 
  $\Gamma = 0.5 \, k_\tf{B} T$, $t_c = 0$, $\lambda = 0$, $\lambda_\tf{S} = 0.5$, and $\lambda_\tf{D} = -0.5$. The chosen trajectory is given by 
  $\bar{E} = 5 \, k_\tf{B} T$, $\bar{\epsilon} = 0$, $\delta_E = 5 \, k_\tf{B}T$, and $\delta_\epsilon = 0.2 \, k_\tf{B}T$.}
  \label{fig:etatb}
\end{figure*}
\begin{figure}[!h]
  \centering
  \includegraphics[width=1.0\columnwidth]{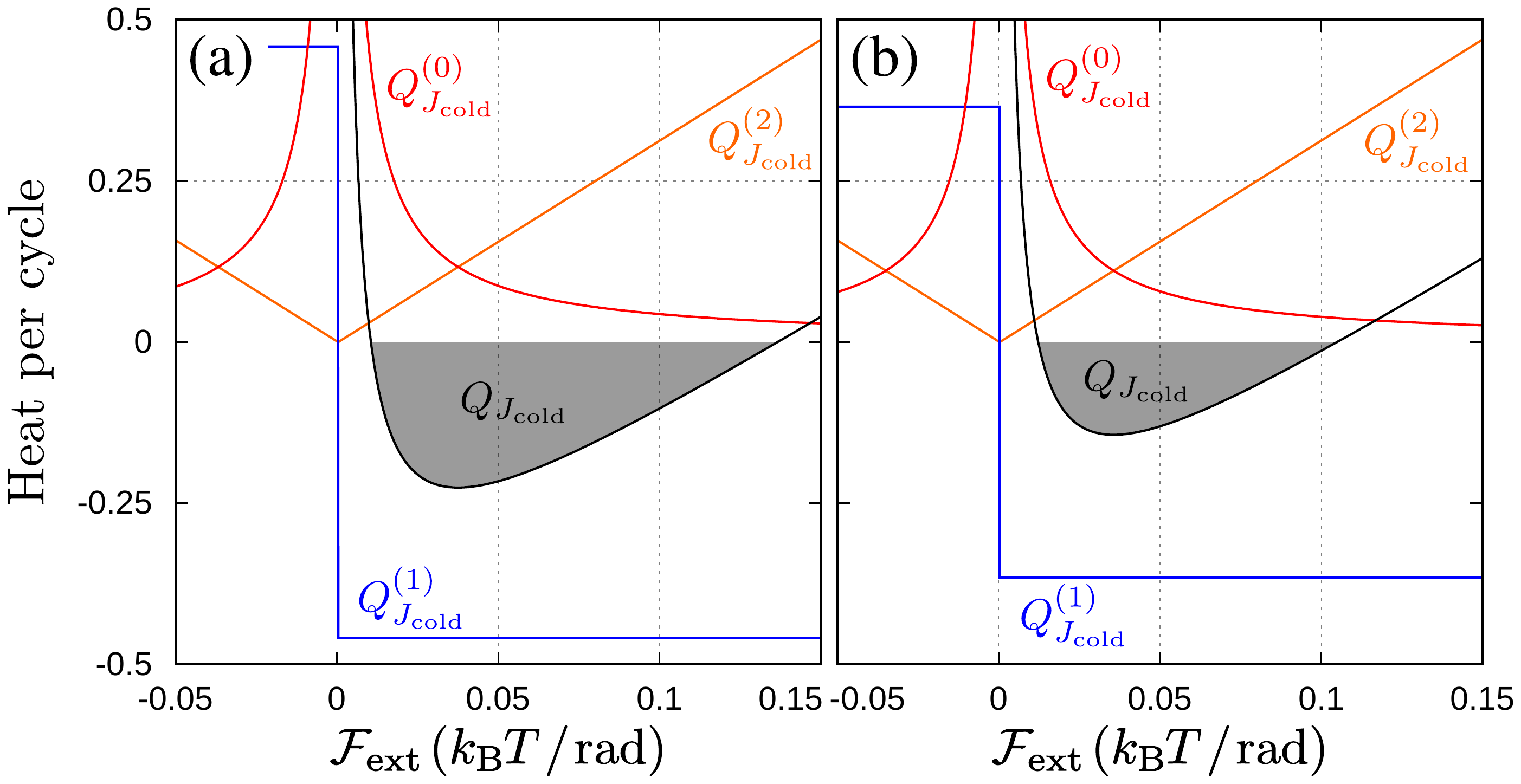}
  \caption{Different order contributions to the transported head from the cold reservoir as a function of the external torque $\mathcal{F}_\tf{ext}$. 	The sum of all these contributions, denoted by $Q_{J_\tf{cold}}$, is shown in solid black. The gray area indicates the region where the device is   
  capable of pumping heat and therefore becomes operational. We considered the same parameters as in Fig.~\ref{fig:etatb}(c) with $\Gamma_\phi = 0$ 
  (a) and $\Gamma_\phi = 0.05 \Gamma$ (b).}
\label{fig:heat}
\end{figure}

In Fig.~\ref{fig:etavb}(a) we show a map of the electric motor/pump efficiencies $\eta_\tf{em}$ and $\eta_\tf{ep}$, see Eqs.~(\ref{eq:eta_em}) and (\ref{eq:eta_ep}), as a function of the external torque $\mathcal{F}_\tf{ext}$ and the bias voltage $\Delta V$ for the decoupled parallel configuration.
The plot allows us to visualize the regions in which the device becomes operational, while giving us a quantitative idea on the performance it can reach. The vertical arms of this cross-shaped map correspond to the motor efficiency [see also panel (b), which shows $\eta_\tf{em}$ for $\Delta V = 0.3 \, k_\tf{B}T$], while the horizontal arms depict the charge pump efficiency [see also panel (c), which shows $\eta_\tf{ep}$ for $\Delta V = 0.03 \, k_\tf{B}T$].
The figure shows that the studied device can indeed act as a motor or a pump. However, the operational regions are very limited in the parameters' space; see the colored regions.
The operational regions of the motor (vertical arms) depend on $\Delta V$ (as a rule of thumb the greater the voltage the greater the CIFs) and on $\mathcal{F}_\tf{ext}$ (at some point this quantity becomes so large that it can not be counteracted by the CIFs).
As mentioned before, the operational regions of the pump (horizontal arms) are subjected to the balance between the different components $Q_I^{(k)}$ of the total transported charge per period.
Finally, it should be mentioned that not all regions in the shown map are consistent with the adiabaticity condition discussed in Sec.~\ref{subsec:mastereq} upon which our expansion is justified.
As a guideline, the unshaded regions in the figure correspond to $\Omega \delta \epsilon / \Gamma k_\tf{B} T \leq 0.05$, in which our expansion should be adequate.

In Figs.~\ref{fig:etavb}(b) and \ref{fig:etavb}(c) we show the typical behavior of the motor and pump efficiencies, respectively, as a function of the external force for a fixed bias voltage. For the case of the motor, panel (b), the shape of the efficiency curve can be understood by taking two limiting situations.
When $\mathcal{F}_\mathrm{ext} \sim \mathcal{F}^{(0)}$, the rotor's frequency goes to zero; cf. Eq.~(\ref{eq:dottheta}). In this situation there is a huge waste of energy given by the zeroth-order leakage current, since $Q_I^{(0)} \propto \Omega^{-1}$. Consequently, the denominator of Eq.~(\ref{eq:eta_em}) becomes so large that the motor's efficiency is strongly reduced.
When $\mathcal{F}_\mathrm{ext} \sim 0$, the stationary-state condition of Eq.~(\ref{eq:stationary}) implies that the numerator of Eq.~(\ref{eq:eta_em}) tends to zero, and so the motor's efficiency. This means that most of the work done by the CIF ($\mathcal{W}_F^{(0)}$) is wasted as heat due to the electronic friction ($\mathcal{W}_F^{(1)}$).

For the case of the pump, Fig.~\ref{fig:etavb}(c), there are also two limiting cases where the efficiency gets suppressed due to the leakage currents and their dependence on $\Omega$. On the one hand, at small $\mathcal{F}_\mathrm{ext}$ we have relatively low velocities, so $\Omega$ is small and $Q_I^{(0)}$ becomes dominant. On the other hand, a large $\mathcal{F}_\mathrm{ext}$ produces a large rotor's velocity, such that in this case $Q_I^{(2)}$ is the main source of leakage. In these situations the pumping current, independent of $\mathcal{F}_\tf{ext}$, is not enough as to counteract the leakage currents, so the total transported charge per cycle still flows in the bias direction, given by $I^{(0)}$. This can be seen in Fig.~\ref{fig:charge}(a), where the three contributions to the total transported charge are shown as a function of the external force.

To evaluate the role of decoherence, in Figs.~\ref{fig:etavb}(b) and \ref{fig:etavb}(c) we also show the device's efficiency for different values of $\Gamma_\phi$. In general, we could say that the decoherence rate has an adverse effect over the performance in the sense that it reduces the maximum efficiency. This result was expected since, as seen before in Sec.~\ref{subsec:coherences}, $\Gamma_\phi$ reduces the peaks in the curvature $\mathcal{B}^I$ ($\mathcal{B}^F$), which is proportional to the pumped charge (CIF work).
For the electric motor this is quite evident as all curves progressively fall off below the zero-decoherence case.
Note also that the range in which the device acts as a motor always decreases with $\Gamma_\phi$. This range is determined by the crossing of the curves with the line $\eta = 0$; see Fig.~\ref{fig:etavb}(b).
For the electric pump, however, the situation is different. There, even though there is an overall decrease of the efficiency, decoherence injection makes the system more resilient to the effects of the external force.
In Fig.~\ref{fig:etavb}(c) we can see that for $\Gamma_\phi = 0$, the device can only bear torques up to approximately 0.3 $k_\tf{B}T/\tf{rad}$, a small value compared to the torques it can withstand for $\Gamma_\phi > 0$.
Interestingly, this allows the system to be ``activated'' by decoherence in regions where, in principle, it would not be operational.
Such effect has already been discussed in a similar quantum system in Ref.~[\onlinecite{fernandez2015}], where, for certain parameter conditions, the performance of an adiabatic quantum motor was improved with the aid of decoherence.
In the mentioned reference the device was described within the Landauer-Büttiker formalism, as is usual in systems where the Coulomb interaction between electrons can be taken as a mean field. Here we see that a similar decoherence-induced activation appears in the pumping regime but under strong Coulomb interaction.
The reason for this decoherence-induced activation becomes clearer when comparing Figs.~\ref{fig:charge}(a) and \ref{fig:charge}(b). There, one can notice that, although all contributions to $Q_I$ decrease with an increase of decoherence, the second-order term is affected much more dramatically, making it negligible within the shown range of parameters. As a consequence of that, the range over which charge pumping is possible gets considerably extended.

In the energy range shown in Fig.~\ref{fig:etavb}(a) and for the considered set of parameters, the maximum (motor) efficiency achieved was $\eta_{\max} \approx 0.01$, a small value if one compares it with the strong interdot coupling regime discussed in Ref.~[\onlinecite{calvo2017}], where efficiencies up to 75\% were obtained. While in the motor operation mode this value can be increased (see below), for the electric pump, the small value of $\eta$ (up to $\sim 0.7 \, \eta_\tf{max}$) seems not so easy to overcome.
The reason behind this again lies in the interplay between the different orders of the transported charge per cycle which, as discussed, obey different laws as one moves the external torque.
As a consequence, the first-order contribution, which is the relevant quantity for the device's performance, only surpasses the sum of the other contributions in a small range of the external force, demarcated by the gray area in Fig.~\ref{fig:charge}.
Increasing the voltage only worsens the situation.
As can be seen in Fig.~\ref{fig:etavb}(a), there is only a limited range for $\Delta V$ in which charge can be pumped. Outside this region $Q_I^{(1)}$ cannot exceed the other contributions, independently of the value of $\mathcal{F}_{\mathrm{ext}}$. In other words, for a large $\Delta V$ there is an overlap of the regions where the dominant contribution is either $Q_I^{(0)}$ or $Q_I^{(2)}$.
Summarizing, this limitation in the operational range of the parameters reduces the possibilities to increase the efficiency of the charge pump. 
Despite that, we remark that the goal of the present work is not to perform an exhaustive search for highly efficient pumps, but to study up to what extent coherences play a role in QD-based quantum machines within the Coulomb blockade regime.

As mentioned, the situation is quite different for the motor regime. There, although the operational region is still limited by $\mathcal{F}_\tf{ext}$, this can be compensated by increasing $\Delta V$. Therefore, there is more freedom to explore the space of parameters in this operational mode. Although not shown, this allowed us to find conditions where adiabatic quantum motors can achieve efficiencies up to 50\% (for $\Delta V \sim 20 \, k_\tf{B}T$).

Recently there has been an increasing interest in studying different forms of heat machines, which drove us to address other operational regimes of our system.
In particular, we explored its role as a heat pump (refrigerator) and as a heat engine (temperature-driven motor); see Fig.~\ref{fig:etatb}.
The results are similar to those described above for the charge pump and the electric motor regimes.
The main differences are: (1) the heat pump is more sensitive to decoherence --see Fig.~\ref{fig:etatb}(c) and notice the difference in the value of $\Gamma_\phi$ with respect to Fig.~\ref{fig:etavb}(c)-- and (2) due to the way in which the efficiencies are affected by $\Gamma_\phi$, we can conclude that there is no activation by decoherence at least in this regime of the parameters.
We observe that efficiencies of the order of 4\% were obtained when taking temperature gradients close to the limit of zero temperature in the cold reservoir [where Carnot's efficiency is 1 and which is out of the range of Fig.~\ref{fig:etatb}(a)]. On the other hand, the quantum refrigerator achieves an efficiency which is approximately 2\% of Carnot's limit.
Again, these values are small when compared to the ones reported for the strong interdot coupling regime, where efficiencies higher than 50\% of Carnot's limit were obtained for both the heat engine and the refrigerator operational modes~\cite{juergens2013,bustos2019}. As discussed above, the reason behind these low values lies in the fact that leakage currents are dominant in the considered regime of parameters. In Fig.~\ref{fig:heat} we show the contributions for the transported heat per cycle as a function of the external torque for $\Gamma_\phi = 0$ and 0.05 $\Gamma$. We can see that both the zeroth- and second-order transported heat (i.e., those coming from the leakage currents) are almost not affected by decoherence, while the first order contribution clearly decays with $\Gamma_\phi$.

\section{Conclusions}
\label{sec:conclusions}

We studied quantum-dot-based nanomachines in the Coulomb blockade regime in a situation where the coherences can dominate the transport properties of the device. We focused our analysis on what we called the decoupled parallel configuration. In this setup, coherences do not come from the interdot coupling which is zero, but from the particles entering/leaving the two dots simultaneously. Therefore, the only way particles entering the system get information from the two dots is through a coherent superposition of states. This makes the modulation manifold effectively bi-parametric, as required in the adiabatic regime. In this sense, the decoupled parallel configuration can be used as the basis for different forms of ``true'' quantum machines, namely: quantum motors, quantum pumps, quantum heat engines, and quantum heat pumps.

We analyzed the impact of decoherence on the above machines. As expected, we found that the overall result is to decrease the efficiency of the machines. In the strong decoherence limit, this can be interpreted as the situation in which the quantum superposition is destroyed, so the electrons in the device can no longer access the two parameters, and the amount of pumped charge/heat or useful work per cycle goes to zero. However, for intermediate values of $\Gamma_\phi$, its effect is more complex due to two main factors.
The first one is that, although decoherence tends to decrease the maximum of the geometric curvatures (current, heat, and force), it also widens them. The second factor is that decoherence can affect the various orders of the adiabatic expansion of the observables in a different way. These two factors are the reason for the found differences between charge and heat pumps regarding the effect of decoherence on them. Importantly, this also causes that, under specific parameters, some forms of quantum machines can be activated by decoherence, in the sense that they require a minimum amount of it to operate. Indeed, such a ``decoherence activation'' mechanism appears, provided that coherences are still present in the local system.

\vspace{0.5cm}

\noindent
\textit{Acknowlegdments.--} We acknowledge financial support by Consejo Nacional de Investigaciones Cient\'ificas y T\'ecnicas (CONICET); Secretar\'ia de Ciencia y Tecnolog\'ia de la Universidad Nacional de C\'ordoba (SECYT-UNC); and Agencia Nacional de Promoción Científica y Tecnológica (ANPCyT, PICT-2018-03587).

\appendix
\section{Trajectory in the parameter space}
\label{app:trajectory}

In Sec.~\ref{subsec:physical} we stated that it is convenient to take an elliptic trajectory around the origin of the energy space in order to take advantage of the shape of $\mathcal{B}^I$ and thus increase the efficiency of the device. More specifically, this elliptic trajectory should be much wider along the $E$ axis than in the $\epsilon$ axis [cf. Fig.~\ref{fig:parallelBI}(a)]. With this in mind, we now show how the experimental setup displayed in Fig.~\ref{fig:scheme} can be configured to allow for such a trajectory. Let $C_1$ and $C_2$ be the capacitances of the side contacts and $C_0$ the capacitance of the central contact, displayed in the middle of the two dots.
For the considered configuration, the dots' energies can be described by
\begin{equation}
  E_i(\theta) = E_i^{(0)} + \frac{q_0(\theta)}{C_0} + \frac{q_i(\theta)}{C_i}, \quad i=\{1,2\},
\end{equation}
where $E_i^{(0)}$ is the energy in the absence of contacts, and the $q$'s denote the amount of charge accumulated in each one of the contacts, as a function of the rotor's position. For the specific geometry of the rotor and the used configuration for the contacts, we could argue that $q_1(\theta) = - q_2 (\theta)$, as the charges in the rotor are assumed to be the same in magnitude, but opposite in sign. Besides, we could simplify the above dependence by stating that $C_1 = C_2$. Due to the position of the central contact with respect to the $C_1$ contact, it is reasonable to expect a phase shift of $\pi/2$ in $q_0$, i.e., $q_1(\theta) = q_0(\theta+\pi/2)$. Accordingly, we replace these assumptions in the above expressions and obtain
\begin{equation}
  E_{1,2}(\theta) = E_{1,2}^{(0)}+\frac{q_0(\theta)}{C_0} \pm \frac{q_0(\theta+\pi/2)}{C_1}.
\end{equation}
If we now define $E=(E_1+E_2)/2$, $\epsilon=E_1-E_2$, $E^{(0)}=(E_1^{(0)}+E_2^{(0)})/2$, $\epsilon^{(0)}=E_1^{(0)}-E_2^{(0)}$, $C_E = C_0$ and $C_\epsilon=C_1/2$, we arrive at the following parametric equations:
\begin{equation}
  E(\theta) = E^{(0)} + \frac{q_0(\theta)}{C_E}, \quad
  \epsilon(\theta) = \epsilon^{(0)} + \frac{q_0(\theta+\pi/2)}{C_\epsilon}.
\end{equation}
Thus, if an elliptic trajectory with $E_\tf{max} \gg \epsilon_\tf{max}$ is desired, then it is enough to take $C_\epsilon \gg C_E$.

\section{Effective evolution kernel}
\label{app:kernel}
In this appendix we show how the blocks of the effective evolution kernel $W^\tf{eff}$ are related to the energy parameters $E$ and $\epsilon$. As discussed in Sec.~\ref{subsec:mastereq}, this effective kernel is defined as the sum of the evolution kernel $W$ and the Liouvillian $L$, which can be decomposed into two contributions, $L_\tf{dot}$ and $L_c$, by separating the $t_c$-dependent term in the electronic Hamiltonian of Eq.~(\ref{eq:helddot}). To study the energy dependence, we will treat these components individually. For the system treated in this work, the matrix elements of the evolution kernel $W$ depend on the DQD's eigenenergies in the following way~\cite{leijnse2008}:

\begin{widetext}
\begin{equation}
	[W]^{a_{0+},a_{0-}}_{a_{2+},a_{2-}} = \tf{i} \sum_{p_2 p_1} \sum_{r\eta} \sum_{a_{1+},a_{1-}} p_2 p_1
	\left[ \sum_\sigma \Pi^{a_{2 p_2} a_{1 p_2}}_{r\sigma(\bar{\eta}p_2)} \Pi^{a_{1 p_1} a_{0 p_1}}_{r\sigma(\eta p_1)} \right]
	\delta_{a_{2\bar{p}_2},a_{1\bar{p}_2}} \delta_{a_{1\bar{p}_1},a_{0\bar{p}_1}}
	\left[ p_1 \phi_r(q^{r,\eta}_{a_{1+},a_{1-}})+\tf{i}\pi f(p_1 q^{r,\eta}_{a_{1+},a_{1-}}) \right],
	\label{eq:kernel_app}
\end{equation}
\end{widetext}
where
\begin{eqnarray}
\Pi^{aa'}_{r\sigma+} &=& \sqrt{\rho_r} \sum_\ell t_{r\ell}      \bra{a} \hat{d}_{\ell\sigma}^\dag \ket{a'}, \notag \\
\Pi^{aa'}_{r\sigma-} &=& \sqrt{\rho_r} \sum_\ell t_{r\ell}^\ast \bra{a} \hat{d}_{\ell\sigma}      \ket{a'}. \label{eq:tme_app}
\end{eqnarray}
Here, $p_i = \pm$ is an index that distinguishes forward ($+$) from backward ($-$) time evolutions on a Keldysh double contour diagram while $\eta = \pm$ is a particle index denoting the annihilation/creation of an electron in the $r$ lead. We use the shorthand notation $\bar{p}_i = -p_i$ and $\bar{\eta} = -\eta$. The indexes $a_{i+}$ and $a_{i-}$ run over the DQD's eigenstates, and $f(x)= [1+\exp(x)]^{-1}$ is the usual Fermi function. The function $\phi(x)$ is defined as
\begin{equation}
\phi_r(x) = - \tf{Re} \, \psi \! \left(\frac{1}{2} + \tf{i} \frac{x}{2\pi} \right) + \ln \frac{D}{2\pi k_\tf{B}T_r},
\end{equation}
where $\psi$ is the digamma function and $D$ denotes the reservoir's bandwidth, which we assume to be independent of $r$ for simplicity. The $q$ argument in the above functions corresponds to the energy difference between initial and final eigenstates, with respect to the $r$-lead electrochemical potential and divided by the thermal energy, i.e.,
\begin{equation}
q_{a_{1+},a_{1-}}^{r,\eta} = \frac{E_{a_{1+}}-E_{a_{1-}}-\eta\mu_r}{k_\tf{B}T_r}.
\end{equation}
If we set the energy of the empty state as reference, i.e., $E_0 = 0$, then all nonvanishing elements of $W$ depend only on the energies of the singly occupied states $\ket{\ell \sigma}$. At the same time, as in the approximation mentioned in Sec.~\ref{subsec:regime} the effective kernel $W^\tf{eff}$ must be taken up to first order in $\Gamma$, and since all elements in $W$ are multiplied by a prefactor proportional to $\Gamma$, the energy differences entering in $q$ need to be taken up to zeroth order in the small parameters $\epsilon \sim t_c \sim \Gamma$. This means that the $q$ argument can only retain the zeroth-order contribution, so all elements in $W$ only depend on the mean level energy $E$~\cite{wunsch2005,riwar2010}.

If we now consider the Liouvillian $L_c$, we can see that
\begin{equation}
  [L_c]_{a,b}^{a',b'}=\bra{a}\hat{H}_c\ket{a'}\delta_{b,b'}-\bra{b'}\hat{H}_c\ket{b}\delta_{a,a'},
\end{equation}
where $\hat{H}_c$ accounts for the interdot coupling Hamiltonian [last term on the right hand side of Eq.~(\ref{eq:helddot})]. Due to the Kronecker deltas and the off-diagonal structure of $\hat{H}_c$ in the local basis, this Liouvillian will only contribute to the $\tf{dn}$ and $\tf{nd}$ blocks of $W^\tf{eff}$ with terms of the form $\pm \tf{i} \, t_c/2$. Lastly, we study the energy dependence of the Liouvillian $L_\tf{dot}$. In this case it can be shown that
\begin{equation}
  [L_\tf{dot}]_{a,b}^{a',b'} =  (E_a-E_b)\delta_{a,a'}\delta_{b,b'}.
\end{equation}
This means that $L_\tf{dot}$ will only contribute to the $\tf{nn}$ block of $W^\tf{eff}$. In the explicit matrix representation of these superoperators [cf. Eq.~(\ref{eq:kernel})] and since we are working in the local basis, this implies that $\bm{L}_\tf{dot}$ is diagonal, whose elements are $\pm \epsilon$.
The matrix representation of the complete Liouvillian $L = L_\tf{dot}+ L_c$ is thus given by: $\bm{L}_\tf{dd} = \bm{0}$, 
\begin{equation}
\bm{L}_\tf{dn} = \frac{t_c}{2} \left(\begin{array}{cccc}
 0 &  0 &  0 &  0\\
 1 &  0 & -1 &  0\\
 0 &  1 &  0 & -1\\
-1 &  0 &  1 &  0\\
 0 & -1 &  0 &  1
\end{array}\right),
\end{equation}
together with $\bm{L}_\tf{nd}=\bm{L}_\tf{dn}^\tf{T}$ and $\bm{L}_\tf{nn}= \epsilon \, \tf{diag}(1,1,-1,-1)$. Regarding the decoherence rates, all the blocks of the decoherence matrix $\bm{\Gamma}_\phi$ are zero, except for the $\tf{nn}$ block, which is simply $\Gamma_\phi$ times the 4$\times$4 identity matrix.

To summarize this analysis, we conclude that all blocks of $\bm{W}^{\rm{eff}}$ are $E$ dependent but only its $\rm{nn}$ block depends on the energy difference $\epsilon$. With regard to the interdot coupling, we can see that for the configuration in series, the elements in $\bm{L}_c$ are the only ones connecting the $\tf{dd}$ and $\tf{nn}$ blocks of the effective kernel, such that coherences are completely determined by this parameter. However, in the configuration in parallel, additional matrix elements proportional to $t_{r,1}t_{r,2}^*$ [cf. Eqs.~(\ref{eq:kernel_app}) and (\ref{eq:tme_app})] contribute in the $\tf{dn}$ and $\tf{nd}$ blocks of $\bm{W}$, such that coherences may even survive without any interdot coupling.

\textit{Occupations and coherences }-- Now that we know all matrix elements of $\bm{W}^\tf{eff}$, we can solve Eq.~(\ref{eq:W^eff}) for the adiabatic occupation probabilities $\bm{p}_\tf{d}^{(0)}$. The first step is to exclude the zero eigenvalue present in the equation. To this end, we make use of the normalization condition $\bm{e}^\tf{T} \bm{p}^{(0)} = 1$ by introducing the pseudo kernel $\tilde{\bm{W}}$, defined below Eq.~(\ref{eq:pk}). This leads to the matrix equation
\begin{equation}
\left(\begin{array}{cc}
 \tilde{\bm{W}}_\tf{dd} & \tilde{\bm{W}}_\tf{dn} \\
 \tilde{\bm{W}}_\tf{nd} & \tilde{\bm{W}}_\tf{nn} 
\end{array}\right)
\left(\begin{array}{c}
 \bm{p}_\tf{d}^{(0)} \\
 \bm{p}_\tf{n}^{(0)} 
\end{array}\right) = 
\left(\begin{array}{c}
 \bm{v}_\tf{d} \\
 \bm{0} 
\end{array}\right),
\label{eq:kin_eff}
\end{equation}
where $[\bm{v}_\tf{d}]_i = -[\bm{W}^\tf{eff}]_{ii}$ and $i$ belongs to the $\tf{d}$ block. This equation can be solved for both the occupations and the coherences by taking the inverse of the pseudo kernel, i.e.,
\begin{equation}
\bm{p}^{(0)} = \tilde{\bm{W}}^{-1} \bm{v},
\end{equation}
with $\bm{v} = (\bm{v}_\tf{d}, \bm{0})^\tf{T}$. With this solution, one can calculate the nonadiabatic corrections $\bm{p}^{(k)}$ through Eq.~(\ref{eq:pk}). If we now want to know about the contributions to the occupations due to the coherences, at least at zeroth order in $\Omega$, we can return to the $\tf{d}$ block of Eq.~(\ref{eq:kin_eff}), such that
\begin{equation}
 \bm{p}_\tf{d}^{(0)} = [\tilde{\bm{W}}_\tf{dd}]^{-1} \bm{v}_\tf{d} - [\tilde{\bm{W}}_\tf{dd}]^{-1} \tilde{\bm{W}}_\tf{dn} \bm{p}_\tf{n}^{(0)}.
\end{equation}
Interestingly, from this equation we can identify two contributions to the occupation probabilities: (1) the first term, which corresponds to the \textit{incoherent} occupations that would come from a sequential tunneling regime where the coupling to the coherences is neglected, and (2) the second term, associated with the presence of coherent superpositions of the dots' states.

\section{Symmetric couplings to the leads}
\label{app:symmetry}

Here we go into detail on the particular case where the DQD system is coupled to the leads through the same asymmetry factors, i.e., $\lambda_\tf{S} = \lambda_\tf{D}$. Under this condition, the tunneling rates satisfy
\begin{equation}
  \Gamma_{r,i} = \frac{\Gamma_r}{\Gamma_{r'}} \Gamma_{r',i},
\end{equation}
where $r,r'=\{\tf{S},\tf{D}\}$, and $i = \{1,2\}$. In the absence of any bias voltage or temperature gradient, as the asymmetry factors only enter in 
$\bm{W}^\tf{eff}$ through the tunneling rates, this results in similar relations when decomposing the effective kernel in its $r$-lead components, such that the same can be applied for the charge currents [cf. Eq.~(\ref{eq:WIr})],
\begin{equation}
  \bm{W}^\tf{eff}_r = \frac{\Gamma_r}{\Gamma_{r'}} \bm{W}^\tf{eff}_{r'} \quad \Rightarrow \quad I_r^{(k)} =
  \frac{\Gamma_r}{\Gamma_{r'}} I_{r'}^{(k)} . \label{currents}
\end{equation}
On the other hand, charge conservation on the first-order currents gives rise to the following relation:
\begin{equation}
  \sum_r I_r^{(1)} = - \td{}{t} \braket{\hat{n}}^{(0)},
\end{equation}
where $\hat{n}$ is the DQD occupation number operator. Hence, if we make use of Eq.~(\ref{currents}) we get
\begin{equation}
  I_r^{(1)} = - \frac{\Gamma_r}{\Gamma} \td{}{t} \braket{\hat{n}}^{(0)}.
\end{equation}
As the above quantity corresponds to a \textit{total} time derivative of a periodic function, the pumped charge, defined as the time integral of $I_r^{(1)}$ over a period of the modulation cycle, clearly integrates to zero. This result is expected since no net charge can be accumulated/lost in the DQD system after each cycle.

\bibliographystyle{apsrev4-1_title}
\bibliography{cite}

\end{document}